\documentclass[11pt,a4paper]{article}
\usepackage[utf8x]{inputenc}
\usepackage[T1]{fontenc}
\usepackage{mathptmx} 

\usepackage[margin=25truemm]{geometry}
\usepackage[pdftex]{graphicx} 
\usepackage{calc} 
\usepackage{hyperref}
\hypersetup{colorlinks,linkcolor=blue}
\usepackage{bm}
\usepackage{enumitem} 
\usepackage{braket,mathtools,amsmath,amssymb,subcaption}
\usepackage{authblk}
\usepackage[all]{nowidow} 
\usepackage{url}
\usepackage{lipsum} 
\newcommand{\vv}[2]{(v^{x}_{#1},v^{y}_{#2})}
\newcommand{\mbf}[1]{\bm{{#1}}}
\newcommand{\vy}[1]{v^y_{#1}}
\newcommand{\dg}[2]{\text{deg}^{{#1}}({#2})}
\hypersetup{ 	
pdfsubject = {},
pdftitle = {},
pdfauthor = {},
	colorlinks=true,       
	linkcolor=blue,        
	citecolor=magenta,        
	filecolor=magenta,     
	urlcolor=blue       
}

\begin{document} 
\title{
Symmetric higher rank topological phases on generic graphs }

\author{Hiromi Ebisu}
\affil{Department of Physics and Astronomy,
Rutgers University, Piscataway, New Jersey 08854,
USA}
\maketitle
\begin{abstract}
Motivated by recent interests in fracton topological phases, we explore the interplay between gapped 2D~$\mathbb{Z}_N$ topological phases which admit fractional excitations with restricted mobility
and geometry of the lattice on which such phases are placed. 
We investigate the properties of the phases in a new geometric context -- graph theory. By placing the phases on a 2D lattice consisting of two arbitrary connected graphs, $G_x\boxtimes G_y$, we study the behavior of fractional excitations of the phases. We derive the formula of the ground state degeneracy of the phases, which depends on invariant factors of the Laplacian.
 
\end{abstract}

\section{Introduction}
The importance of the discovery of topologically ordered phases can hardly be overstated~\cite{Tsui,Laughlin1983,Kalmeyer1987,wen1989chiral,Wen:1989iv,PhysRevLett.66.1773}. They provide a paradigm shift in understanding phase transitions away from one based purely on symmetry breaking. Topologically ordered phases also admit exotic phenomena, such as fractionalized quasiparticle excitations (i.e., anyons)~\cite{Laughlin1983,leinaas1977theory} and topologically protected ground state degeneracy, independent of the local geometry of the system~\cite{Elitzur:1989nr}. 
These phases also have a great advantage 
for the purposes of quantum computing, as operation on a state in a subspace of degenerate vacua, realized by braiding anyons, is immune to local perturbations~\cite{KITAEV20032,dennis2002topological}.
Theoretical frameworks to describe these phases have been well developed, such as the topological quantum field theory~\cite{witten1989quantum,Elitzur:1989nr,wen2004quantum}
and the modular tensor category~\cite{Kitaev2006}.
\par
Recently, new types of topological phases have been proposed, 
which are beyond these frameworks, 
often called fracton topological phases~\cite{chamon,Haah2011,Vijay}. A unique feature of these phases is that they exhibit the sub-extensive 
ground state degeneracy~(GSD) dependence. Due to the UV/IR mixing propriety, one cannot have effective field theory description in the long wavelength limit. 
The key insight to understand such unusual GSD dependence is that mobility of the quasiparticle excitations is sensitive to the local geometry of the system, which is contrasted with conventional topologically ordered phases where the properties of the excitations depend only on the global topology of the system. Therefore, fracton topological phases hold value for exploring new geometric phases. A theoretical formalism of these phases has yet to be completed. \par 
Due to the sensitiveness of the UV physics in fracton topological phases, it would be interesting to study the phases on a curved geometry. Indeed, several works studied gapless theory with fractonic-like mobility constraint on a curved geometry~\cite{pena2021fractons,bidussi2022fractons,jain2022fractons}, and gapped fracton topological phases on generic lattices~\cite{PhysRevB.97.165106,PhysRevX.8.031051,tian2020haah,manoj2021arboreal}.

In this paper, we introduce unusual gapped~$\mathbb{Z}_N$ topological phases where deconfined fractional excitations are subject to the mobility constraint in a similar fashion as the fracton topological phases and explore the geometric properties of the fractional excitations by placing the phases on the generic lattices beyond the typical square one. In particular, we highlight the behavior of the fractional excitations of the phases in a new geometric context~--~\textit{graph theory}. 
(There have been a few attempts at tackling the problem in this direction, see, e.g, \cite{Fradkin2015,graph2022,gorantla2022gapped}.)
Introducing a 2D lattice composed of an arbitrary two connected graphs, we study the behavior of the excitations and the superselection sectors (i.e., distinct types of excitations) of the model on this lattice. By making use of formalism of the graph theory, one can systematically study the properties of the excitations. 
As we will see in the later section, the properties of the fractional excitations are determined by the Laplacian matrix (the Laplacian in short), which is the graph theoretical analog of the second-order spatial derivative. \par
The Laplacian plays a pivotal role in the graph theory. For instance, one can study the connectivity of the graph by evaluating eigenvalues of the Laplacian~\cite{chung1997spectral}. In our context, the fusion rules of the fractional excitations follows from the form of the Laplacian of the graph, and that the GSD depends on the $N$ and the invariant factors of the Laplacian. Our study might contribute to a better understanding of the fracton topological phases in view of graph theory. 
\par
The outline of this paper is as follows. In Sec.~\ref{sc2}, we introduce the model Hamiltonian. We demonstrate that our simple model of the topological phase is obtained by gapping the gauge group from $U(1)$ to~$\mathbb{Z}_N$ via Higgs mechanism in the unusual Maxwell theory. After obtaining the Hamiltonian, in Sec.~\ref{main}, we consider placing the phase on the 2D lattice constructed of the product of two arbitrary graphs. Section~\ref{sec4} is devoted to elucidating the properties of fractional excitations of the model and identifying GSD. We show that the fusion rules of the fractional excitations are determined by the form of the Laplacian of the graph and that the superselection sectors are associated with the kernel and cokernel (the Picard group) of the Laplacian. We further show that the GSD depends on the invariant factors of the Laplacian. In Sec.~\ref{sec5}, we give a simple example of the lattice to see how our result works. Physical intuition of our result is also given. Finally, in Sec.~\ref{sec6}, we conclude our work with a few future research directions.  


\section{Model Hamiltonian}\label{sc2}
In this section, we introduce the model Hamiltonian. For the sake of clearer illustration, we first focus on the Hamiltonian on the flat space. 
The key insight to obtain the model is gapping the gauge group from $U(1)$ to $\mathbb{Z}_N$ via Higgs mechanism in the unusual Maxwell theory, referred to as the higher rank Maxwell theory in this paper, where the kinetic and potential terms are described by the second-order spatial derivative of the gauge potential, instead of the first order. Accordingly, we dub the phases obtained by this procedure as \textit{higher rank topological phases}.
This procedure is contrasted with the case where the $\mathbb{Z}_N$ topological phase~(toric code) is obtained from the conventional Maxwell theory via Higgs mechanism. See~\cite{PhysRevB.96.125151,PhysRevB.97.235112,Higgs2018,PhysRevB.106.045145,PhysRevB.105.045128} for more explanations on other types of higher rank Maxwell theories and their Higgs phases.
\subsection{Higher rank Maxwell theory}
Before going into the details of the model Hamiltonian, it is useful to discuss the $U(1)$ higher-rank Maxwell theory in the continuum limit. The difference between this theory and the usual Maxwell theory is that the first order spatial derivative operator, which enters in the Gauss law or gauge invariant operators in the conventional Maxwell theory, is replaced with the second-order derivative. 
We start by introducing~$U(1)$ gauge fields in 2D, $A^k(\mathbf{x})$, $E^k(\mathbf{x})$ ($k=x,y$, $\mathbf{x}$: spatial coordinate), which are a canonical conjugate pair:
\begin{equation}
    [A^k(\mathbf{x}),E^l(\mathbf{y})]=i\delta_{k,l}\delta(\mathbf{x}-\mathbf{y})
\end{equation}
Introducing the charge density operator~$\rho(\mbf{x})$ and the second-order spatial derivative operator, $D_k=\partial_k^2$, the Gauss law is given by
\begin{equation}
    \rho(\mathbf{x})=D_kE^k(\mathbf{x}),\label{gauss1}
\end{equation}
where the repeated indices are summed over. We define magnetic flux which is invariant under the Gauss law~\eqref{gauss1} by 
\begin{equation}
    B(\mathbf{x})=D_xE^y(\mathbf{x})-D_yE^x(\mathbf{x}).\label{B}
\end{equation}
An interesting property of this theory is that not only charge but also dipole and quadrupole moments are conserved, which is in contrast to the conventional Maxwell theory where only the charge is conserved. To see how, transform the dipole moment $\int d^2\mbf{x}(x\rho)$ as
\begin{equation}
    \int d^2\mbf{x}(x\rho)= \int d^2\mbf{x}(xD_kE^k(\mathbf{x}))=(\text{boundary term})+ \int d^2\mbf{x}(\partial_x^2(x)E^x)=(\text{boundary term}).\label{dipole}
\end{equation}
Here we have referred to~\eqref{gauss1} and implemented the partial integration twice, yielding only the boundary term (which is constant). Likewise, one can show that $  \int d^2\mbf{x}\rho$,   $\int d^2\mbf{x}(y\rho)$, and $  \int d^2\mbf{x}(xy\rho)$ , corresponding to charge, dipole and quadrupole, are conserved. As we see later, depending on the geometry, conservation of these moments corresponds to the conservation of dipole and quadrupole moments of the fractional excitations in the higher rank $\mathbb{Z}_N$ topological phase. \par

\begin{figure}
    \begin{center}
     \begin{subfigure}[h]{0.42\textwidth}
    \includegraphics[width=\textwidth]{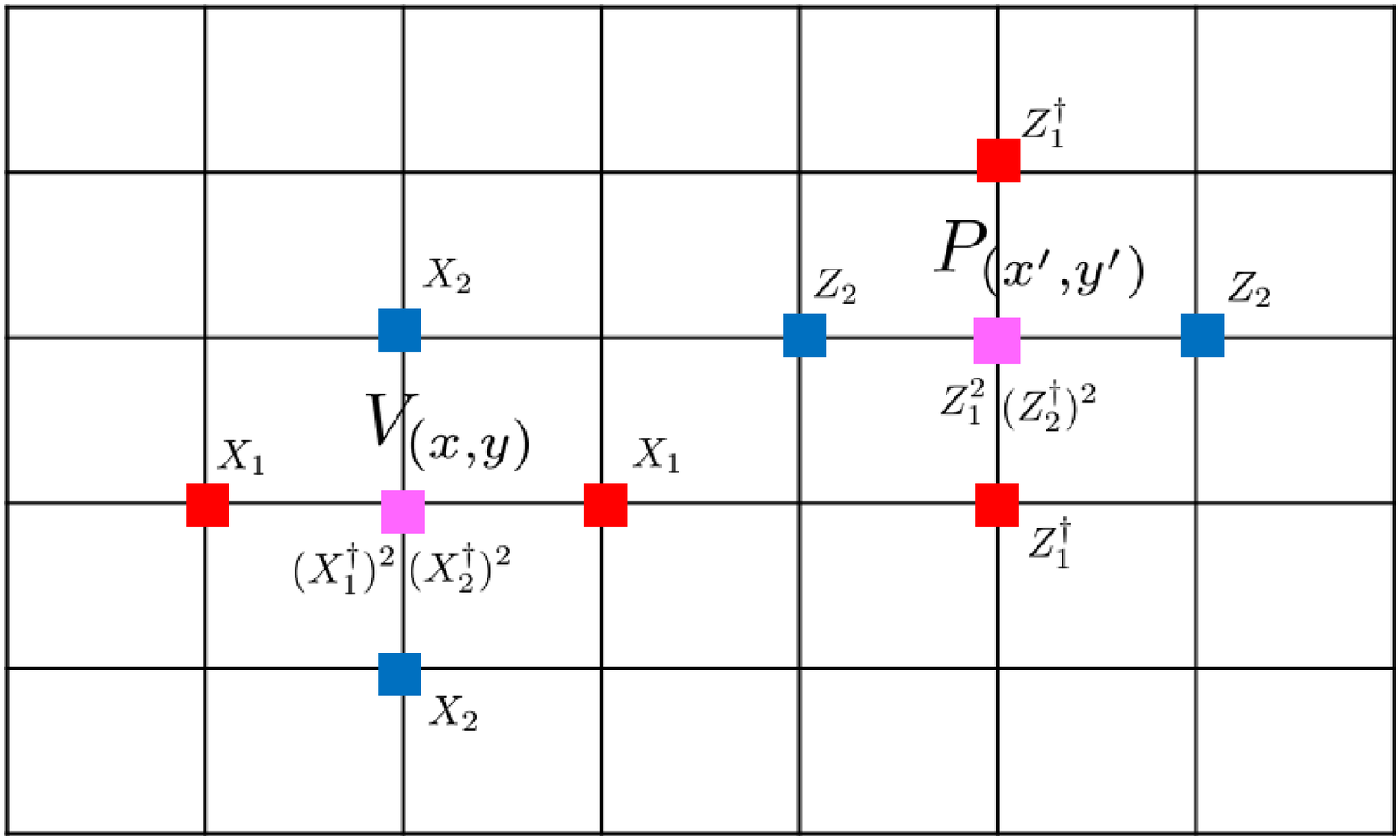}
         \caption{}\label{rule3}
             \end{subfigure} 
        \begin{subfigure}[h]{0.29\textwidth}
    \includegraphics[width=\textwidth]{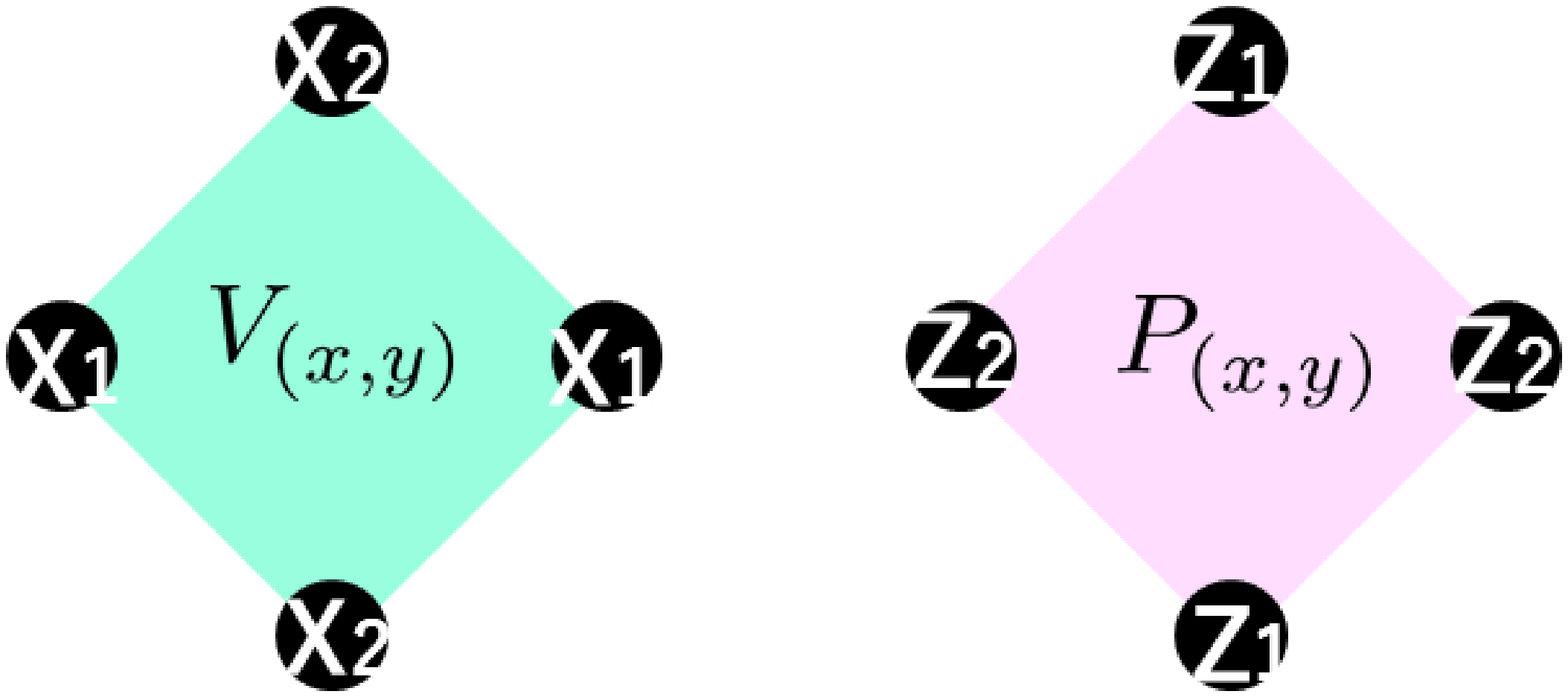}
         \caption{}\label{sf1}
             \end{subfigure}

                    \begin{subfigure}[h]{0.40\textwidth}
    \includegraphics[width=\textwidth]{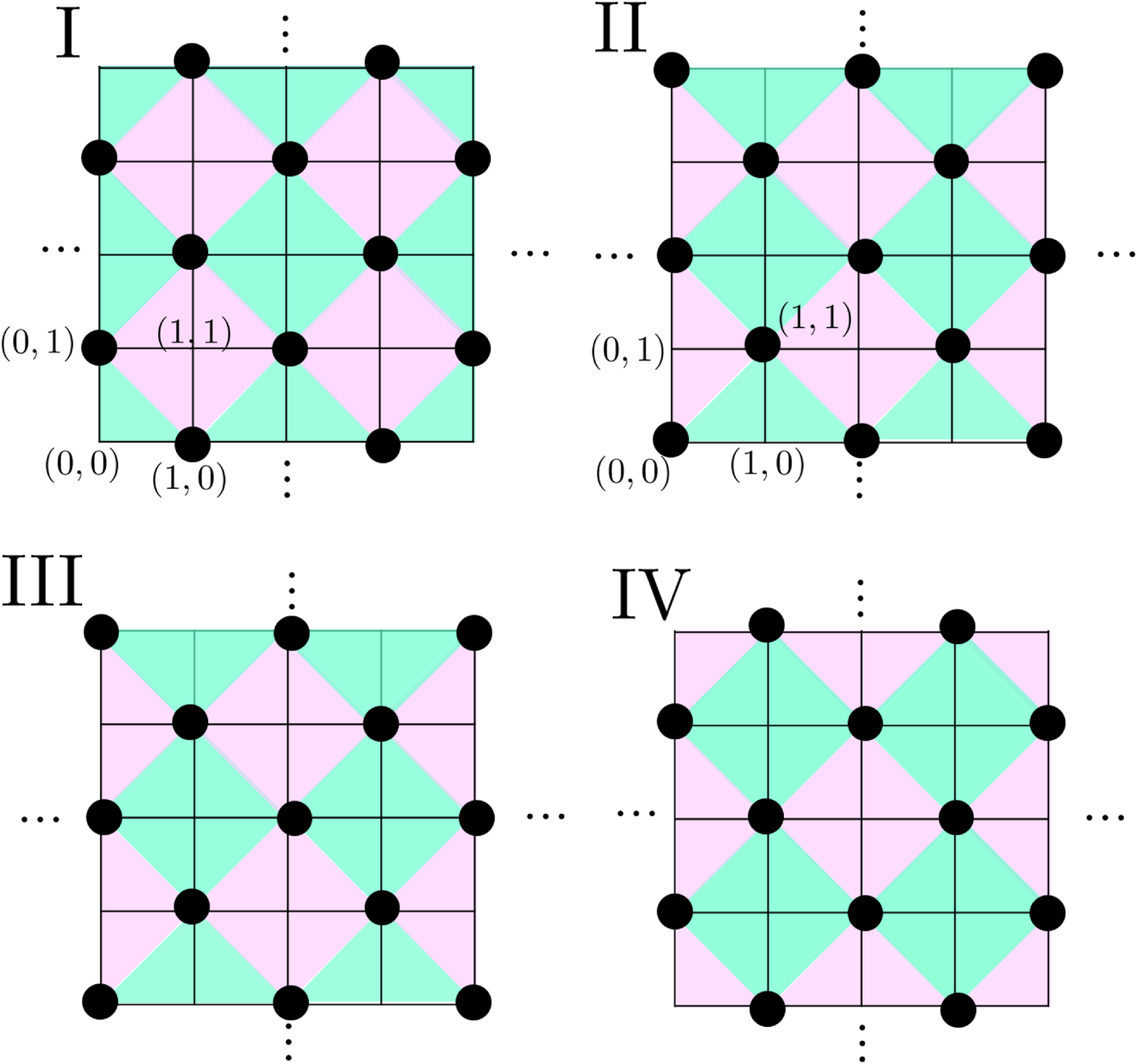}
         \caption{}\label{sf2}
             \end{subfigure} 
                         \begin{subfigure}[h]{0.40\textwidth}
    \includegraphics[width=\textwidth]{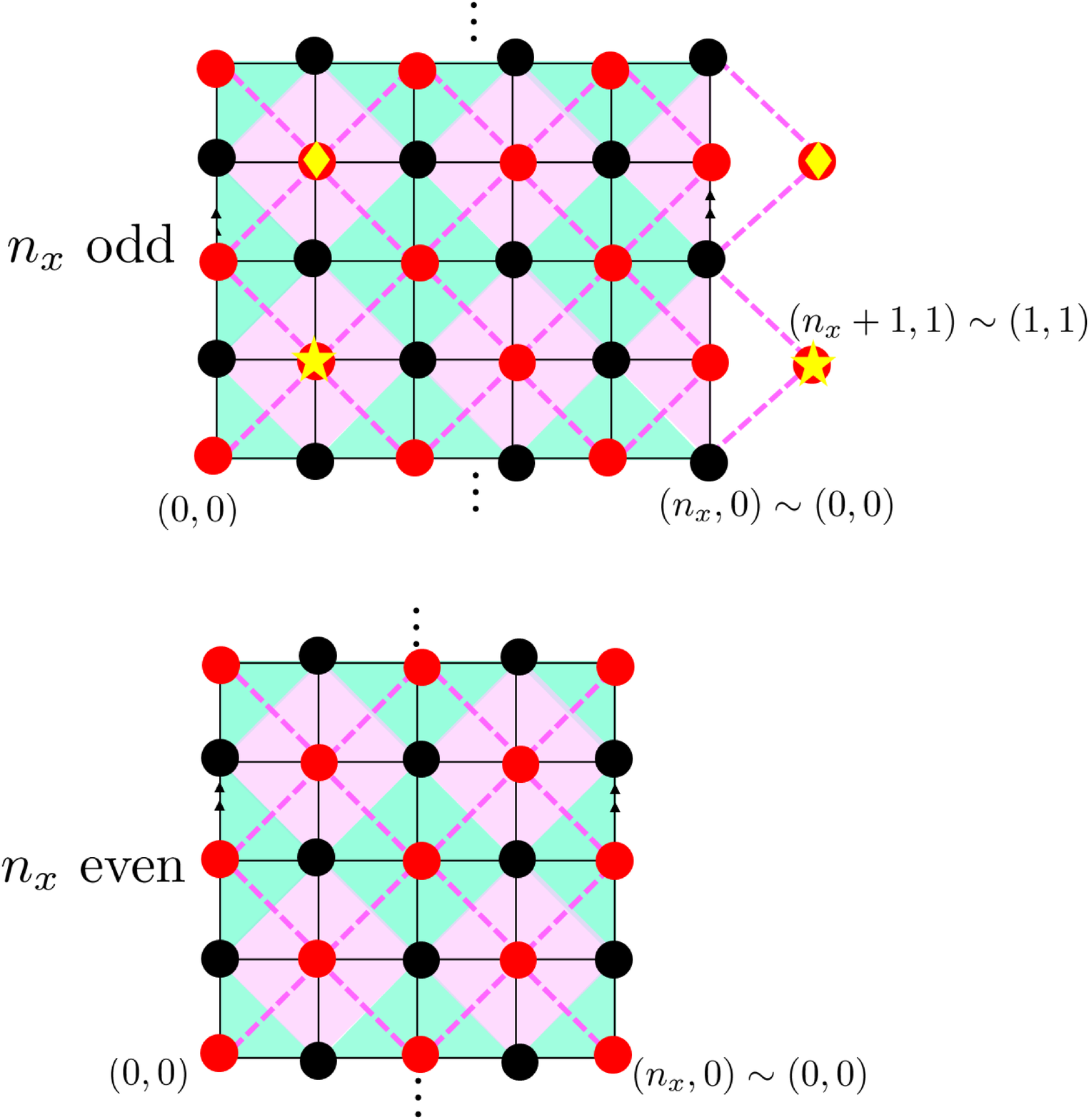}
         \caption{}\label{sf3}
             \end{subfigure} 
                         
             \caption{(a)~Two terms defined in~\eqref{VP} on 
             2D square lattice. (b) These two terms has the simple form~\eqref{VP2} in the case of $N=2$, each of which resembles the ones defined in the $\mathbb{Z}_2$ toric code.
       (c) Configuration of the mutually commuting terms belonging to I-IV defined in~\eqref{four}. (d) Configuration of the terms which belong to I and the one of $V_{(x,y)}$ belonging to II (pink dashed lines and red dots), in the case where periodic boundary condition is imposed with $(n_x,n_y)=$(odd, even)[top] and the one with $(n_x,n_y)=$(even, even)[bottom]. For illustration purposes, we slightly extend the geometry, identifying the vertices
       with the same symbols (yellow star and rhombus) due to the periodic boundary condition.}
 \end{center}
 \end{figure}

Now we place this theory on the 2D square lattice and gap it to $\mathbb{Z}_N$ via Higgs mechanism, which can be accomplished by two steps. First, discretize the spatial coordinate $\mbf{x}$ by introducing the lattice coordinate so that $\mbf{x}\to(x,y)\in \lambda(\mathbb
{Z},\mathbb{Z})$ with $\lambda$ being lattice spacing. 
The two pairs of gauge potential and electric field, which are canonical conjugate, are now labeled by $(A^k_{(x,y)},E^l_{(x,y)})$ with relation
\begin{equation*}
    [A^k_{(x,y)},E^l_{(x^\prime,y^\prime)}]=i\delta_{k,l}\delta_{x,x^\prime}\delta_{y,y^\prime}.
\end{equation*}
We transform the second-order spatial derivative into the discretized form ($D_k\to\nabla_{k}^2$). 
The Gauss law~\eqref{gauss1} becomes
\begin{equation}
    \rho_{(x,y)}=\nabla^2_xE^x_{(x,y)}+\nabla^2_yE^y_{(x,y)}=
    (E_{(x+1,y)}^x+E_{(x-1,y)}^x-2E_{(x,y)}^x)+(E_{(x,y+1)}^y+E_{(x,y-1)}^y-2E_{(x,y)}^y)\label{gauss}.
\end{equation}
Similarly, the magnetic flux operator, corresponding to~\eqref{B}, is defined as
\begin{equation}
    B_{(x,y)}=\nabla^2_xA^y_{(x,y)}-\nabla^2_yA^x_{(x,y)}=
    (A_{(x+1,y)}^y+A_{(x-1,y)}^y-2A_{(x,y)}^y)-(A_{(x,y+1)}^x+A_{(x,y-1)}^x-2A_{(x,y)}^x).\label{flux}
\end{equation}
The second step is condensing charge~$N$ excitations, reducing the $U(1)$ gauge group down to $\mathbb{Z}_N$. As a consequence, the gauge fields take $\mathbb{Z}_N$ value: $A_{(x,y)}^{k}=\frac{2\pi\mathbb{Z}}{N}$ (mod $2\pi\mathbb{Z}$). 
The gauge and electric fields are expressed via 
\begin{equation}
    Z_{1,(x,y)}=e^{iA^x_{(x,y)}}, \;X_{1,(x,y)}=\omega^{E^x_{(x,y)}},Z_{2,(x,y)}=e^{iA^y_{(x,y)}}, \;X_{2,(x,y)}=\omega^{E^y_{(x,y)}}, \label{a1}
\end{equation}
where $\omega$ denotes the $N$th root of unity, i.e., $\omega=e^{i2\pi/N}$. 
These operators act on the local $N\times N$ dimensional Hilbert space~$\ket{a}_{(x,y)}\ket{b}_{(x,y)}\;(a,b\in\mathbb{Z}_N)$ as
\begin{eqnarray}
Z_{1,(x,y)}\ket{a}_{(x,y)}\ket{b}_{(x,y)}=\omega^a\ket{a}_{(x,y)}\ket{b}_{(x,y)},\;Z_{2,(x,y)}\ket{a}_{(x,y)}\ket{b}_{(x,y)}=\omega^b\ket{a}_{(x,y)}\ket{b}_{(x,y)}\nonumber\\
X_{1,(x,y)}\ket{a}_{(x,y)}\ket{b}_{(x,y)}=\ket{a+1}_{(x,y)}\ket{b}_{(x,y)},\;X_{2,(x,y)}\ket{a}_{(x,y)}\ket{b}_{(x,y)}=\ket{a}_{(x,y)}\ket{b+1}_{(x,y)}
\end{eqnarray}
indicating that \eqref{a1} represents the~$\mathbb{Z}_N$ Pauli algebra. 
From the expression~\eqref{a1}, we can define the $\mathbb{Z}_N$ Gauss and flux operator as (see Fig.~\ref{rule3})
\begin{eqnarray}
V_{(x,y)}\equiv\omega^{\rho_{(x,y)}}=X_{1,(x+1,y)}X_{1,(x-1,y)}(X_{1,(x,y)}^{\dagger})^2X_{2,(x,y+1)}X_{2,(x,y-1)}(X_{2,(x,y)}^{\dagger})^2,\nonumber\\
P_{(x,y)}\equiv e^{iB_{(x,y)}}=Z_{1,(x,y+1)}^{\dagger}Z_{1,(x,y-1)}^{\dagger}Z^2_{1,(x,y)}Z_{2,(x+1,y)}Z_{2,(x-1,y)}(Z_{2,(x,y)}^{\dagger})^2.\label{VP}
\end{eqnarray}
By construction, these two operators commute. It is important to note that the form of the operators~\eqref{VP} is determined by the discretized second-order derivative, $\nabla_k^2$. 
The Hamiltonian of the $\mathbb{Z}_N$ Higgs phase whose ground state is a state without charge and flux, is defined by
\begin{equation}
    H_{\mathbb{Z}_N}=-\sum_{x,y}(V_{(x,y)}+P_{(x,y)})+h.c\label{zn}.
\end{equation}
This model shares several features as the
toric code~\cite{KITAEV20032}, in that the ground state~$\ket{\Omega}$ is the stabilized state satisfying $V_{(x,y)}\ket{\Omega}=P_{(x,y)}\ket{\Omega}=\ket{\Omega}$, Also, the model admits the two types of deconfined excitations, carrying $\mathbb{Z}_N$ electric and magnetic charges. 
However, there is a crucial difference between our model and the toric code. There is a mobility constraint on the fractional excitations, yielding unusual GSD dependence on the lattice.
\par
\subsection{The simplest example: $N=2$ on the square lattice -- decoupled toric codes}\label{2pt2}
To get a handle on the physical intuition behind the Hamiltonian~\eqref{zn}, and see how the GSD of the model drastically changes depending on the lattice, 
it is useful to take a closer look at the model in the simplest case by setting $N=2$ on the square lattice before considering the phases on generic lattices constructed by graphs. For a moment, we consider 2D square lattice without boundary. 
In the case of $N=2$, the two terms~\eqref{VP} are simplified (Fig.~\ref{sf1}):
\begin{eqnarray}
V_{(x,y)}=X_{1,(x+1,y)}X_{1,(x-1,y)}X_{2,(x,y+1)}X_{2,(x,y-1)}\nonumber\\
P_{(x,y)}=Z_{1,(x,y+1)}Z_{1,(x,y-1)}Z_{2,(x+1,y)}Z_{2,(x-1,y)}.\label{VP2}
\end{eqnarray}
The Hamiltonian~\eqref{zn} with~\eqref{VP2} resembles the $\mathbb{Z}_2$ toric code~\cite{KITAEV20032} with a crucial difference that the terms~$V_{(x,y)}$ and $P_{(x,y)}$ involve four \textit{next}-nearest neighboring Pauli operators in the horizontal and vertical direction, not nearest neighbors. 
Due to this property, one can classify the mutually commuting terms~\eqref{VP2} into the following four groups:
\begin{eqnarray}
   && \text{I}:\{V_{(2m,2n)},P_{(2m^\prime-1,2n^\prime-1)}\}\;\; \text{II}:\{V_{(2m-1,2n)},P_{(2m^\prime,2n^\prime-1)}\}\nonumber\\
   && \text{III}:\{V_{(2m,2n-1)},P_{(2m^\prime-1,2n^\prime)}\}\;\; \text{IV}:\{V_{(2m-1,2n-1)},P_{(2m^\prime,2n^\prime)}\}\;\;(m,n,m^\prime,n^\prime \in\mathbb{Z}).\label{four}
\end{eqnarray}
We portray these configurations of the terms in Fig.~\ref{sf2}, which are reminiscent of the ones found in the~$\mathbb{Z}_2$ surface code~\cite{surfacecode2012}.\par

Now we impose the boundary condition on the lattice and evaluate the GSD. Suppose we impose the periodic boundary condition with lattice length, $n_x$, $n_y$,  being even number of sites in both of the $x$ and $y$-directions, which is schematically described by $(n_x,n_y)=(\text{even, even})$. In this case, the Hamiltonian~\eqref{zn} with~\eqref{VP2} can be decomposed into four according to~\eqref{four}, i.e., the Hamiltonian consists of four decoupled~$\mathbb{Z}_2$ toric codes. Since the GSD of each $\mathbb{Z}_2$ toric code on torus is given by $4$, the GSD of the model is found to be $4^4=256$.
The situation differs when the length of the lattice is set to be odd. For instance, when the length of the lattice in the $x$-direction is odd while keeping the one in the $y$-direction being even, i.e., $(n_x,n_y)=(\text{odd, even})$, one cannot separate the terms belonging to I and II as well as III and IV. Indeed, the terms which belong to I are ``connected" with the ones belonging to II.
For instance, as demonstrated in the top geometry in Fig.~\ref{sf3}, the terms $P_{(n_x-2,2n^\prime-1)}$ which belong to I and $P_{(n_x+1,2n^\prime-1)}$ belonging to II are located adjacent with each other, which is opposed to the case with $n_x$ being even where $P_{(n_x-2,2n^\prime-1)}$ and $P_{(n_x+1,2n^\prime-1)}$ are decoupled (bottom geometry in Fig.~\ref{sf3}). A similar argument holds for the terms $V_{(x,y)}$. Analogous lines of thought leads to that one cannot separate terms belonging to III and IV.
Therefore, the mutually commuting terms fall into two groups:
\begin{equation*}
\text{I}^\prime:\{V_{(m,2n)},P_{(m^\prime,2n^\prime-1)}\}\;\; \text{III}^\prime:\{V_{(m,2n-1)},P_{(m^\prime,2n^\prime)}\},
\end{equation*}
implying that we have two decoupled $\mathbb{Z}_2$ toric codes. Thus, the GSD is given by $4^2=16$. One can similarly discuss the GSD in other cases of the length of the lattice. Overall, we have 
\begin{equation}
    GSD=\begin{cases}
        256\;\bigl[(n_x,n_y)=(\text{even, even})\bigr]\\
        16\;\bigl[(n_x,n_y)=(\text{odd, even}), (\text{even, odd})\bigr]\\
         4\;\bigl[(n_x,n_y)=(\text{odd, odd})\bigr].
    \end{cases}
   \end{equation}
 To summarize this subsection, in the simplest case, we learn that each term which constitutes the Hamiltonian involves the next-nearest neighbors corresponding to the second-order derivative of the higher rank Maxwell theory and, due to this property, 
 the GSD drastically changes depending on whether the length of the lattice is even and odd. As we will see in the later section, this feature can be understood in terms of graph theory. Indeed, the GSD depends on $N$ and the invariant factors of the Laplacian.
 
\section{Putting the theory on graphs}\label{main}
 In this section, we introduce a lattice consisting of two arbitrary connected graphs and place the model Hamiltonian~\eqref{zn} on it. The central idea is that when placing the Hamiltonian~\eqref{zn} on a graph, we 
replace the derivative operators $\nabla^2_k$ defined on the square lattice with the \textit{Laplacian}, which is the graph theoretical analog of the second-order derivative~\cite{chung1997spectral}. In accordance with this replacement, the Gauss law and the flux operator given in \eqref{VP} is modified. 
\par
\subsection{Notations from graph theory}
Let us first give a formal definition of a graph $G=(V,E)$. It is a pair
consisting of a set of vertices~$V$ and a set of edges $E$ composed of pairs of vertices~$\{v_i,v_j\}$. Throughout this paper, we assume that the graph is \textit{connected}, i.e., there is a path from a vertex to any other vertex (there is no isolated vertex), and that the graph does not have an edge that emanates from and terminates at the same vertex.
We also define
two quantities, deg$(v_i)$ and $l_{ij}$, which play pivotal roles in this paper. The former one, deg$(v_i)$ denotes the \textit{degree} of the vertex $v_i$, i.e., the number
of edges emanating from the vertex $v_i$ and the latter one, $l_{ij}$ represents the number of edges between two vertices $v_i$ and $v_j$~(we have $l_{ij}=0$ when there is no edge between two vertices, $v_i$ and $v_j$.).
Using these two quantities, 
the \textit{Laplacian matrix} of the graph is defined.
For a given graph $G=(V,E)$, the Laplacian matrix $L$ (which we abbreviate as Laplacian in the rest of this work) is the matrix with rows and columns indexed by the elements of vertices~$\{v_i\}\in V$, with
\begin{equation}
    L_{ij}=\begin{cases}\text{deg}(v_i)\;(i=j)
    \\-l_{ij}\;(i\neq j)
    \end{cases}.\label{laplacian}
\end{equation}
The Laplacian is singular 
due to the connectivity of the graph. (Summing over all rows or columns gives zero.)
As an example, the Laplacian of the cycle graph $C_3$ (i.e., a triangle) consisting of three vertices and three edges, where there is a single edge between a pair of vertices, is given by
\begin{equation*}
    L=\begin{pmatrix}
2 & -1&-1 \\
-1 & 2&-1 \\
-1&-1&2
\end{pmatrix}.
\end{equation*}\par

\subsection{2D lattice and Hamiltonian}
With these preparations, now we introduce the 2D lattice. 
Let $G_x(V_x,E_x)$ and $G_y(V_y,E_y)$ be two connected graphs. We denote vertices of these two graphs as $v_i^{x}$ and $v_j^{y}$ ($1\leq i\leq n_x$, $1\leq j\leq n_y$), where $n_x (n_y)$ represents the total number of vertices in graph $G_x (G_y)$. 
Moreover, the Laplacian of the graph $G_x(G)_y$ is denoted as $L_x(L_y)$ whose matrix elements are defined by~\eqref{laplacian}, i.e., 
the Laplacian $L_x$ is defined by
\begin{equation*}
    (L_x)_{i,i^\prime}
    =\begin{cases}\text{deg}^x(v^x_i)\;(i=i^\prime)
    \\-l^x_{ii^\prime}\;(i\neq i^\prime)
    \end{cases}\;\;(1\leq i,i^\prime\leq n_x), 
\end{equation*}
and the Laplacian $L_y$ is similarly introduced.\par
\begin{figure}
    \begin{center}
        \begin{subfigure}[h]{0.52\textwidth}
    \includegraphics[width=\textwidth]{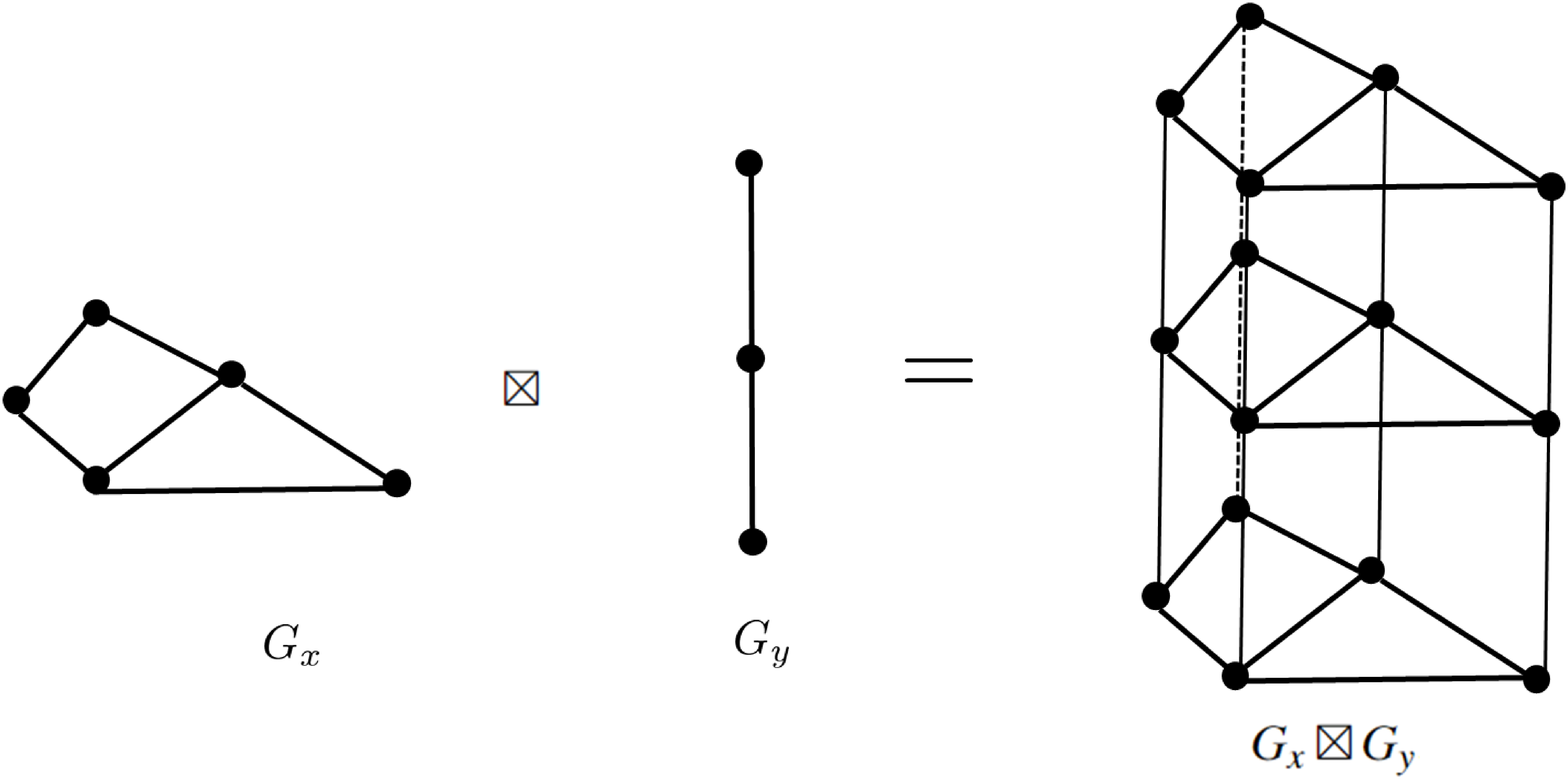}
         \caption{}\label{rule2}
             \end{subfigure} 
               \begin{subfigure}[h]{0.42\textwidth}
    \includegraphics[width=\textwidth]{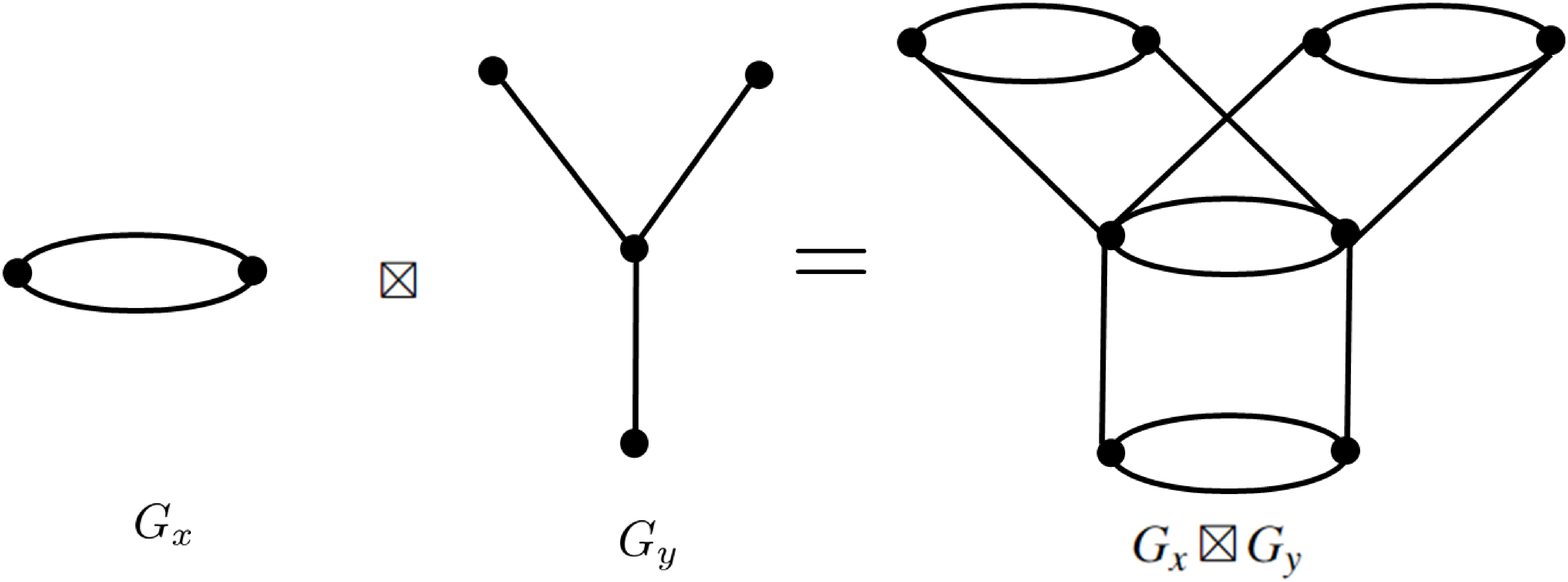}
         \caption{}\label{rule20}
             \end{subfigure} 
                    \begin{subfigure}[h]{0.45\textwidth}
    \includegraphics[width=\textwidth]{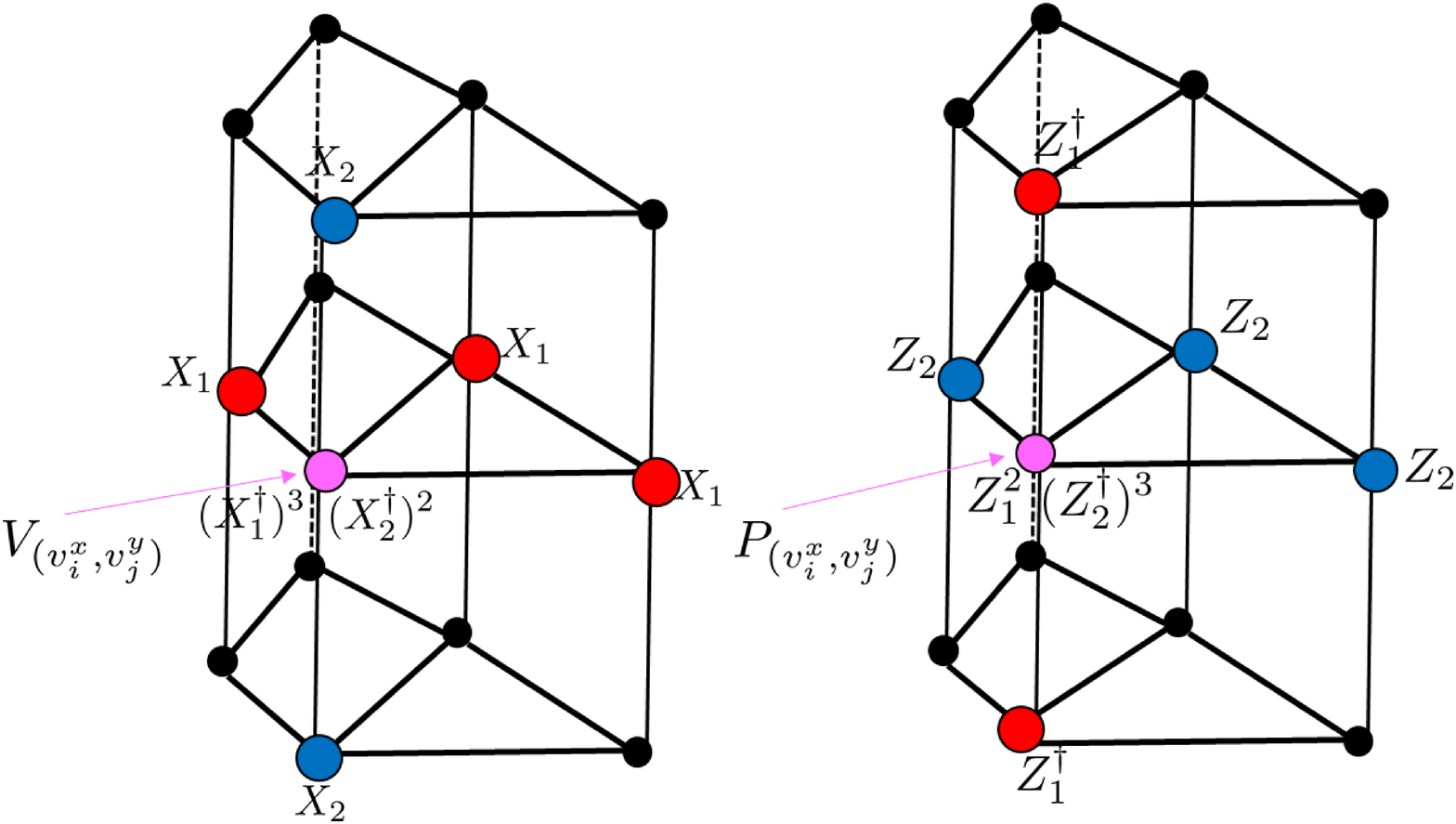}
         \caption{}\label{rule5}
             \end{subfigure} 
                         
             \caption{(a)(b)Two examples of the 2D lattice comprised of two connected graphs, $G_x\boxtimes G_y$. (c) Two terms given in~\eqref{term2} which are defined on the lattice $G_x\boxtimes G_y$ given in (a).  }
 \end{center}
 \end{figure}

The 2D lattice is introduced by the product of the two graphs,~$G_x\boxtimes G_y$, where each coordinate of the vertex is represented by $\vv{i}{j}$. Intuitively, the lattice is constructed by ``stacking the graph $G_x$ along the graph $G_y$", meaning, the graph $G_x$ is attached 
at each vertex of the graph $G_y$, $\vy{j}$ and how these $G_x$'s are connected follows from edges of the graph $G_y$. 
We portray examples of such lattices in Fig.~\ref{rule2}~\ref{rule20}.
(Note that the lattice considered here is defined on an abstract 2D cell complex. Indeed, each graph
consists of vertices and edges, corresponding to zero- and one-simplices. Due to this fact, we regard $G_x\boxtimes G_y$ as the 2D lattice.) The square lattice (without taking into the account the boundary) can be reproduced by setting $\text{deg}^x(v^x_i)=\text{deg}^y(v_j^y)=2$, $l^x_{i,i^\prime}=\delta_{i,i^\prime\pm1}$, $l^y_{j,j^\prime}=\delta_{j,j^\prime\pm1}$.
\par
We place the higher rank $\mathbb{Z}_N$ topological phase on this lattice $G_x\boxtimes G_y$ by defining the $U(1)$ higher rank Maxwell theory on the graph and gapping the gauge group to~$\mathbb{Z}_N$ similarly to the case of the square lattice presented in the previous section.
Since the procedure closely parallels the one in the previous section except that we define the second-order derivative via the Laplacian $L_x$ and $L_y$, we outline the procedure succinctly. In the 2D lattice $G_x\boxtimes G_y$, we introduce two pairs of the $U(1)$ gauge potential and electric field, which are canonical conjugate, $(A^k_{\vv{i}{j}},E^k_{\vv{i}{j}})$ acting on the coordinate $\vv{i}{j}$
with relation
\begin{equation*}
    [A^k_{\vv{i}{j}},E^l_{\vv{i^\prime}{j^\prime}}]=i\delta_{k,l}\delta_{i,i^\prime}\delta_{j,j^\prime}.
\end{equation*}

Replacing $\nabla_k^2$ with $-L_k$,
the Gauss law and magnetic flux is defined by
\begin{eqnarray}
    \rho_{\vv{i}{j}}&=&-L_xE^x_{\vv{i}{j}}-L_yE^y_{\vv{i}{j}}\nonumber\\
  B_{\vv{i}{j}}&=&-L_xA^y_{\vv{i}{j}}+L_yA^x_{\vv{i}{j}}.\label{flux2}
\end{eqnarray}
We gap the gauge group from $U(1)$ to down to $\mathbb{Z}_N$ via Higgs mechanism.
Introducing two types of generalized $\mathbb{Z}_N
$ qubit states ($\mathbb{Z}_N$ clock states) on each vertex of the 2D lattice, labeled by 
$\ket{a}_{\vv{i}{j}}\ket{b}_{\vv{i}{j}}$ ($a,b\in\mathbb{Z}_N$), we define the operators acting on these qubits as 
\begin{equation}
    Z_{1,\vv{i}{j}}=e^{iA^x_{\vv{i}{j}}}, \;X_{1,\vv{i}{j}}=\omega^{E^x_{\vv{i}{j}}},Z_{2,\vv{i}{j}}=e^{iA^y_{\vv{i}{j}}}, \;X_{2,\vv{i}{j}}=\omega^{E^y_{\vv{i}{j}}}. \label{a2}
\end{equation}
Analogously to~\eqref{a1}, they form the $\mathbb{Z}_N$ algebra. 
Similarly to~\eqref{VP}, 
we define the $\mathbb{Z}_N$ Gauss and flux terms at each vertex $\vv{i}{j}$ by
\begin{equation*}
    V_{\vv{i}{j}}=\omega^{\rho_{\vv{i}{j}}},\;P_{\vv{i}{j}}=e^{iB_{\vv{i}{j}}}.
\end{equation*}
Referring to \eqref{laplacian} and~\eqref{flux2}, one can rewrite these terms as
\begin{eqnarray}
V_{\vv{i}{j}}&=&\biggl(X^\dagger_{1,\vv{i}{j}}\biggr)^{\dg{x}{v_i^x}}\prod_{s\neq i}X_{1,\vv{s}{i}}^{l^{x}_{sj}}\times\biggl(X^\dagger_{2,\vv{i}{j}}\biggr)^{\dg{y}{v^y_j}}\prod_{t\neq j} X_{2,\vv{t}{j}}^{l^{y}_{tj}}\nonumber\\
    P_{\vv{i}{j}}&=&\biggl(Z^\dagger_{2,\vv{i}{j}}\biggr)^{\dg{x}{v_i^x}}\prod_{s\neq i}Z_{2,\vv{s}{j}}^{l^{x}_{si}}\times Z^{\dg{y}{v_j^y}}_{1,\vv{i}{j}}\prod_{t\neq j} \biggl(Z^{\dagger}_{1,\vv{i}{t}}\biggr)^{l^{y}_{tj}}\label{term2}.
\end{eqnarray}
We portray these terms in Figs.~\ref{rule5} in the same 2D lattice as Fig.~\ref{rule2}.

It is straightforward to check every term given in~\eqref{term2} commute with one another. 
Using these mutual commuting terms, we introduce the Hamiltonian by
\begin{equation}
    H=-\sum_{i,j}V_{\vv{i}{j}}-\sum_{i,j}P_{\vv{i}{j}}+h.c.\label{hamiltonian}
\end{equation}
The ground state is the stabilized state satisfying $V_{\vv{i}{j}}\ket{\Omega}=P_{\vv{i}{j}}\ket{\Omega}=\ket{\Omega}$. In the next section, we discuss the properties of the excitations.

\section{Superselection sectors}\label{sec4}
Now we come to the main part of this paper.
In this section, we discuss the properties of the excitations of the model on the graphs defined in Sec.~\ref{main}.
\subsection{Fusion rules}
Similarly to the toric code, there are two types of excitations of our model, carrying $\mathbb{Z}_N$ electric and magnetic charges, which violates the condition $V_{\vv{i}{j}}\ket{\Omega}=\ket{\Omega}$ and $P_{\vv{i}{j}}\ket{\Omega}=\ket{\Omega}$, respectively. We label these two excitations at coordinate $\vv{i}{j}$, whose eigenvalue of $V_{\vv{i}{j}}$ and $P_{\vv{i}{j}}$ is $\omega$, 
by $e_{\vv{i}{j}}$ and $m_{\vv{i}{j}}$. Also, we label their conjugate with eigenvalue $\omega^{-1}$ by $\overline{e}_{\vv{i}{j}}$ and $\overline{m}_{\vv{i}{j}}$. \par
One can systematically discuss the fusion rules of these fractional excitations.
 \begin{figure}
    \begin{center}
         
       \includegraphics[width=0.35\textwidth]{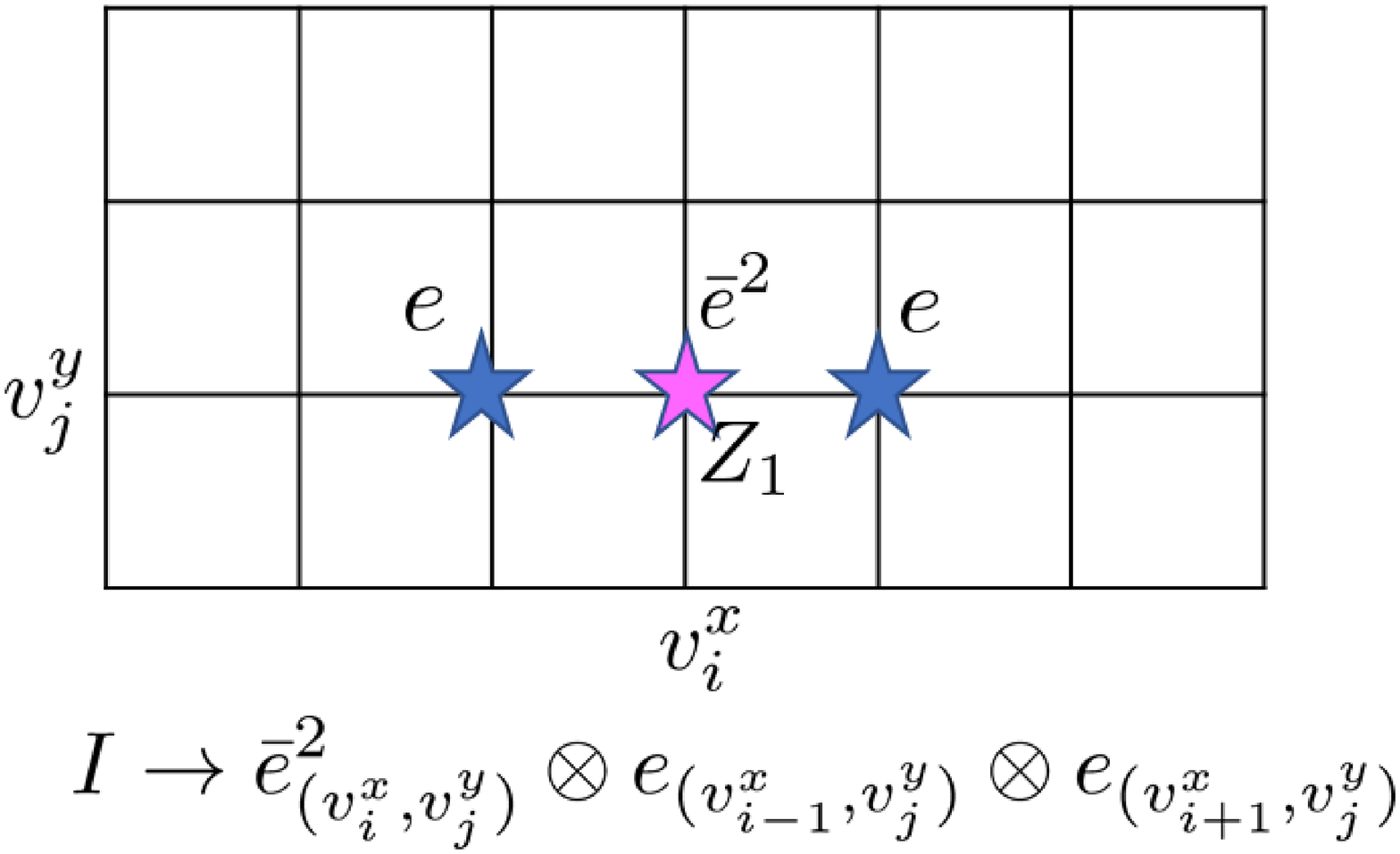}

 \end{center}
       
      \caption{ The fusion rule of the electric charges in the case of the square lattice (without boundary). When a single operator $Z_{1,\vv{i}{j}}$ acts on the ground state, the condition of~$V_{\vv{i}{j}}=1$ is violated at the vertex with coordinate~$\vv{i}{j}$ (pink star) and the adjacent ones, i.e., the ones with~$\vv{i\pm1}{j}$ (blue star), giving excitations. The fusion rule induced by the applying the operator $Z_{1,\vv{i}{j}}$ is schematically described by $I\to \bar{e}^2_{(v^x_i,v^y_j)}\otimes e_{(v^x_{i-1},v^y_j)}\otimes e_{(v^x_{i+1},v^y_j)}$.
 }
        \label{fr}
   \end{figure}
Let us focus on the fusion rules of the electric charges. Applying the $\mathbb{Z}_N$ operator $Z_{1,\vv{i}{j}}$ on the ground state at the coordinate $\vv{i}{j}$, it violates the condition of~$V_{\vv{i}{j}}=1$ at the vertex with coordinate~$\vv{i}{j}$ and the ones connected with edges in the horizontal direction, namely, 
\begin{equation*}
    V_{\vv{i}{j}}(Z_{1,\vv{i}{j}}\ket{\Omega})=\omega^{-\text{deg}^x(v_i)}(Z_{1,\vv{i}{j}}\ket{\Omega}),\;\;V_{(v_s^x,\vy{j})}(Z_{1,\vv{i}{j}}\ket{\Omega})=\omega^{l^x_{si}}(Z_{1,\vv{i}{j}}\ket{\Omega})\;\;(s\neq i).
\end{equation*}
The fusion rule is schematically described by (see also Fig.~\ref{fr} for an example)
\begin{equation}
    I\to (\overline{e}_{\vv{i}{j}})^{\text{deg}^x(v_i)}\otimes \prod_{s\neq i}(e_{(v_s^x,v_j^y)})^{l^x_{si}},\label{fusion}
\end{equation}
where $I$ denotes the vacuum sector. Likewise, if we apply $Z_{2,\vv{i}{j}}$ on the ground state, we have fusion rule
\begin{equation}
     I\to (\overline{e}_{\vv{i}{j}})^{\text{deg}^y(v_j)}\otimes \prod_{t\neq j}(e_{(v_i^x,v_t^y)})^{l^y_{jt}}.\label{fusion2}
\end{equation}
The fusion rules~\eqref{fusion}~\eqref{fusion2} are a generalization of the ones in 2D topologically ordered phases where a pair of anyons are created. One can rewrite the fusion rules~\eqref{fusion}~\eqref{fusion2} more succinctly by using the Laplacian. 
On a lattice $G_x\boxtimes G_y$ at given $v^y_j$, 
we define the~$n_x$-dimensional vector
where each entry takes the $\mathbb{Z}_N$ value by
\begin{equation}
    \bm{r}_{v^y_j}=(r_1,r_2,\cdots,r_{n_x})^T\in\mathbb{Z}_N^{n_x} 
    \label{vector}
\end{equation}
from which we introduce multiple
sets of $Z_1$ operators, $Z_{1,\vv{1}{j}}^{r_1}Z_{1,\vv{2}{j}}^{r_2}\cdots Z_{1,\vv{n_x}{j}}^{r_{n_x}}$ acting on the ground state. For the sake of the simplicity, in the following, we omit the subscript of $\bm{r}_{v^y_j}$ on the left-hand side of~\eqref{vector} and write it as $\bm{r}$ 
till the point where it is necessary to mention the $v_j^y$ dependence.\par
Introducing the fundamental basis of vectors $\{\bm{\lambda}_i\}$ as
$\bm{\lambda}_i=(\underbrace{0,\cdots,0}_{i-1},1,\underbrace{0,\cdots,0}_{n_x-i})^T\in \bm{r}$, the fusion rule~\eqref{fusion} is rewritten as 
\begin{equation}
    I\to {e}_{\vv{1}{j}}^{a^x_1}\otimes {e}_{\vv{2}{j}}^{a^x_2}\otimes\cdots\otimes {e}_{\vv{n_x}{j}}^{a^x_{n_x}}\:(a^x_i\in\mathbb{Z}_N),\label{fusion3}
\end{equation}
with 
\begin{equation}
\bm{f}^x_e\vcentcolon  =(a^x_1,a^x_2,\cdots,a^x_{n_x})^T=-L_x\bm{\lambda}_i.\label{tr}
\end{equation}
Note that in the fusion rule~\eqref{fusion3}, charge conservation is satisfied, i.e., $\sum_i a^x_i=0\; (\text{mod} N)$ as the Laplacian $L_x$ is singular (summing over matrix elements along the~$i$th column gives zero). One can similarly describe the fusion rule~\eqref{fusion2} in terms of the Laplacian $L_y$.\par
We can also systematically discuss the fusion rules of the electric charges induced by applying \textit{multiple sets} of $Z_1$ or $Z_2$ operators on the ground state instead of applying a single operator. When we apply $Z_{1,\vv{1}{j}}^{r_1}Z_{1,\vv{2}{j}}^{r_2}\cdots Z_{1,\vv{n_x}{j}}^{r_{n_x}}$ on the ground state, characterized by vector $\bm{r}$~\eqref{vector}, the fusion rule of the electric charges has the same form as~\eqref{fusion3} by setting 
\begin{equation}
    \bm{f}^x_e=-L_x\bm{r}.\label{tr2}
\end{equation}
One can write the fusion rules by applying sets of $Z_2$ operators as well as the ones for magnetic charges in a similar manner. Since discussion of these fusion rules closely parallels what we have just discussed, we do not present it here.  
\par
As we will see in the next subsection, 
the way we describe the fusion rules~\eqref{fusion3}\eqref{tr2} turn out to be useful to
discuss the number of distinct fractional charges in our model on the graph.
\subsection{Ground state degeneracy}
In this subsection, we derive the formula of the GSD of our model on the graph. To this end, 
we count the distinct types of quasiparticle excitations. The spirit behind such counting is analogous to~\cite{graph2022}. In the derivation, we will use the key property of the Laplacian; introducing the invertible integer matrices $P$ and $Q$, the Laplacian can be transformed into the diagonal form (\textit{Smith normal form}) via
\begin{equation}
    PLQ=\text{diag}(u_1,u_2,\cdots,u_{n-1},0)\vcentcolon  = D, \label{snf}
\end{equation}
where $u_i$ represents positive integers, satisfying $u_i|u_{i+1}$ for all $i$ (i.e., $u_i$ divides $u_{i+1}$ for all~$i$)~\cite{lorenzini2008smith}. Since the Laplacian is singular, the last diagonal entry is zero. The diagonal element~$u_i$, referred to as the \textit{invariant factors} of the Laplacian, %
plays a pivotal role in the graph theory. In what follows, we will see the GSD is characterized by these invariant factors of the Laplacian. This can be achieved by two steps. First, we count the number of distinct loops in the horizontal direction. Second, we evaluate the distinct number of configurations of such loops up to deformation in the vertical direction. 
\subsubsection{The number of closed loops in the horizontal direction}\label{hori}
To start, 
we first count the number of distinct loops of electric charges in the horizontal direction, i,e, the number of closed loops of the electric charges at given $v^y_j$. The loop is constructed by a ``string" of the~$Z_1$ operators, $Z_{1,\vv{1}{j}}^{r_1}Z_{1,\vv{2}{j}}^{r_2}\cdots Z_{1,\vv{n_x}{j}}^{r_{n_x}}$ characterized by the vector, $\bm{r}$~\eqref{vector}. The loops must commute with terms $V_{\vv{i}{j}}$ defined in \eqref{term2}, which means the composite of the operators $Z_{1,\vv{1}{j}}^{r_1}Z_{1,\vv{2}{j}}^{r_2}\cdots Z_{1,\vv{n_x}{j}}^{r_{n_x}}$ does not create an excitation. This condition amounts to be that the fusion rule induced by such a product of the operators becomes trivial. Referring to~\eqref{fusion3}\eqref{tr2}, 
such condition is rewritten as 
\begin{equation}
    L_x\bm{r}=\mathbf{0}\mod N.\label{con1}
\end{equation}
Therefore, to count the distinct loops of the electric charges in the horizontal direction, we need to evaluate the kernel of the Laplacian, $L_x$. Note that since the graph is connected, meaning the summing over the entries of the Laplacian along any row gives zero, there are at least $N$ solutions of~\eqref{con1}, $\bm{r}=h(1,1,\cdots,1)^T\;(h\in\mathbb{Z}_N)$. 
\par
To proceed, we transform the Laplacian $L_x$ into the Smith normal form~\eqref{snf}.
Introducing integer matrices $P_x$ and $Q_x$ whose absolute value of the determinant is one, we can transform the Laplacian into the Smith normal form:
\begin{equation}
    P_xL_xQ_x=\text{diag}(u^x_1,\cdots,u^x_{n_x-1},0)\vcentcolon=D_x, 
\end{equation}
from which we have 
\begin{eqnarray}
\eqref{con1}\Leftrightarrow P_x^{-1}D_xQ_x^{-1}\bm{r}=\bm{0}\mod N\nonumber\\\Leftrightarrow
   D_x \tilde{\bm{r}}=\mathbf{0}\mod N. 
   \label{con20}
        \end{eqnarray}
When moving from the second to the third equation, we have used the fact that $P_x$ is the integer matrix, and we have defined 
$\tilde{\bm{r}}\vcentcolon=Q_x^{-1}\bm{r}$. \par
Suppose there are $m_x$ invariant factors of $L_x$ which are greater than one, i.e.,
\begin{equation}
    D_x=\text{diag}(\underbrace{1,\cdots,1}_{n_x-1-m_x},\underbrace{p_1,\cdots,p_{m_x}}_{m_x},0),\label{snf22}
\end{equation}
then, from~\eqref{con20}, it follows that the first $n_x-1-m_x$ components of the vector $\mathbf{\tilde{r}}$ are zero:
\begin{equation}
    \tilde{r}_{a^\prime}=0 \mod N\;(1\leq a^\prime\leq n_x-1-m_x).
\end{equation}
Regarding the elements $\tilde{r}_{a+n_x-1-m_x}$ $(1\leq a\leq m_x)$, one finds
\begin{equation}
  p_a\tilde{r}_{a+n_x-1-m_x}=0\mod N\;\Leftrightarrow   \;p_a\tilde{r}_{a+n_x-1-m_x}= Nt_a\;(1\leq a\leq m_x,\;t_a\in\mathbb{Z}).\label{pr}
\end{equation}
Decompose $N$ and $p_a$ into two integers as 
\begin{equation}
    N=N^{\prime}_i \gcd(N,p_a), \;p_a=p_a^\prime \gcd(N,p_a),\label{nprime}
\end{equation}
where gcd stands for the greatest common divisor and $N^\prime_a$ and $p_a^\prime$ are coprime, \eqref{pr} becomes
\begin{equation*}
    p_a^\prime\tilde{r}_{a+n_x-1-m_x}=N_a^\prime t_a.
\end{equation*}
Since $N^\prime_a$ and $p_a^\prime$ are coprime, one finds
\begin{equation}
    \tilde{r}_{a+n_x-1-m_x}=N^\prime_a \alpha_a\;(1\leq a\leq m_x),
\end{equation}
where integer $\alpha_a$ takes $\gcd(N,p_a)$ distinct values, i.e., $\alpha_a=0,1,\cdots,\gcd(N,p_a)-1$.
There is no constraint on the last element of $\bm{\tilde{r}}$, $\tilde{r}_{n_x}$ as the last diagonal entry of $D_x$ is zero. This implies that $\tilde{r}_{n_x}$ takes $N$ distinct values.\par
Overall, with the assumption of \eqref{snf22}, the condition~\eqref{con20} gives
\begin{equation}
   \bm{\tilde{r}}=(\underbrace{\tilde{r}_1,\cdots,\tilde{r}_{n-1-m_x}}_{n_x-1-m_x},\underbrace{\tilde{r}_{n_x-m_x},\cdots,\tilde{r}_{n_x-1}}_{m_x},\tilde{r}_{n_x})^T=(\underbrace{0,\cdots,0}_
   {n_x-1-m_x},\underbrace{N^\prime_1 \alpha_1,\cdots,N^\prime_{m_x} \alpha_{m_x}}_{m_x},\alpha_{m_x+1})^T \mod N
  ,\label{con3}
\end{equation}
where $0\leq \alpha_a\leq \gcd(N,p_a)-1 (1\leq a\leq m_x)$, $0\leq \alpha_{m_x+1} \leq N-1$.
Thus, the kernel of the Laplacian, which is associated with the closed loops of electric charges, is labeled by 
\begin{equation}
    \mathbb{Z}_{\gcd(N,p_1)}\times \mathbb{Z}_{\gcd(N,p_2)}\times\cdots\times\mathbb{Z}_{\gcd(N,p_{m_x})}\times \mathbb{Z}_N=\prod_a\mathbb
    {Z}_{\gcd(N,p_a)}\times \mathbb{Z}_N.
\label{eee}
\end{equation}
Recalling $\tilde{\bm{r}}\vcentcolon=Q_x^{-1}\bm{r}$, the form of the loop, $\bm{r}$ is obtained by multiplying $Q_x$ from the left in~\eqref{con3}. Writing the~$n_x\times n_x$ matrix $Q_x$ as 
\begin{equation}
    Q_x=(\underbrace{\bm{q}_1,\cdots,\bm{q}_{n_x-1-m_x}}_{n_x-1-m_x},\underbrace{\bm{\tilde{q}}_{1},\cdots,\bm{\tilde{q}}_{m_x}}_{m_x},\bm{\tilde{q}}_{m_x+1}),\label{loop3}
\end{equation}
where each column is given by an $n_x$ dimensional vector, we have 
\begin{eqnarray}
    \bm{r}&=&Q_x\tilde{\bm{r}}=\alpha_1N_1^\prime\bm{\tilde{q}}_1+\cdots+\alpha_{m_x}N_{m_x}^\prime\bm{\tilde{q}}_{m_x}+\alpha_{m_x+1}\bm{\tilde{q}}_{m_x+1}\nonumber\\
    &\vcentcolon=&\alpha_1\bm{\Lambda}_{1}+\cdots+\alpha_{m_x}\bm{\Lambda}_{m_x}+\alpha_{m_x+1}\bm{\Lambda}_{m_x+1}
    .\label{loop51}
\end{eqnarray}

\subsubsection{Deformation of the closed loops -- analogy to the chip-firing game}
After identifying the loops of electric charge in the horizontal direction, we need to count the 
number of distinct configurations of such loops up to the deformation in the $y$-direction by the sets of $P_{(v_x,v_y)}$. 
This feature is contrasted with the toric code, where the non-contractible loop in the horizontal direction is deformed so it is shifted up or downward. In our case, the way of the loops being deformed is not so immediate as the toric code. We will see that to describe the deformation of the loops, the Laplacian comes into play. 
\par
For the sake of the illustration, we focus on the case where the 2D lattice is $C_{n_x}\boxtimes C_{n_y}$ for the moment and then move on to more general cases of the graph later. 
Here,~$C_{p}$ represents the cyclic graph consisting of $p$ vertices in a cyclic order where the adjacent vertices are connected with an edge. 
In particular, we set $N=3$ and consider the case with $C_{6}\boxtimes C_{6}$. 
The coordinate of the lattice is labeled by $\vv{i}{j}\;(1\leq i,j\leq 6)$, where vertex $v_i^x$ ($v^y_j$) is aligned in cyclic order along the horizontal (vertical) direction. This geometry is nothing but the 2D torus.
As explained in more detail in the next section (Sec.~\ref{sec5}), the Smith normal form of the Laplacian of $C_6$ reads
\begin{equation*}
    D_x=\text{diag}(1,1,1,1,6,0),
\end{equation*}
from which the closed loop of the electric charge at $v^y_j$ is labeled by $\mathbb{Z}_3\times\mathbb{Z}_3$ [\eqref{eee}]. Furthermore, by evaluating $Q_x$, and referring to~\eqref{loop51}, the form of the closed loop at $v^y_j$,
$Z_{1,\vv{1}{j}}^{r_1}Z_{1,\vv{2}{j}}^{r_2}\cdots Z_{1,\vv{n_x}{j}}^{r_{n_x}}$ characterized by vector $\bm{r}$,
is found to be 
\begin{equation}
    \bm{r}_{v^y_j}=\alpha_{1,\vy{j}}(2,1,0,2,1,0)^T+\alpha_{2,\vy{j}}(1,1,1,1,1,1)^T\vcentcolon=\alpha_{1,\vy{j}}\bm{\Lambda}_{1,v^y_j}+\alpha_{2,\vy{j}}\bm{\Lambda}_{2,v^y_j},\;\;(\alpha_{1,\vy{j}},\alpha_{2,\vy{j}})\in\mathbb{Z}_3^2
\end{equation}
where we retrieve the subscript, emphasizing 
$v^y_j$ dependence. 
\begin{figure}
    \begin{center}
       \begin{subfigure}[h]{0.85\textwidth}
  \includegraphics[width=\textwidth]{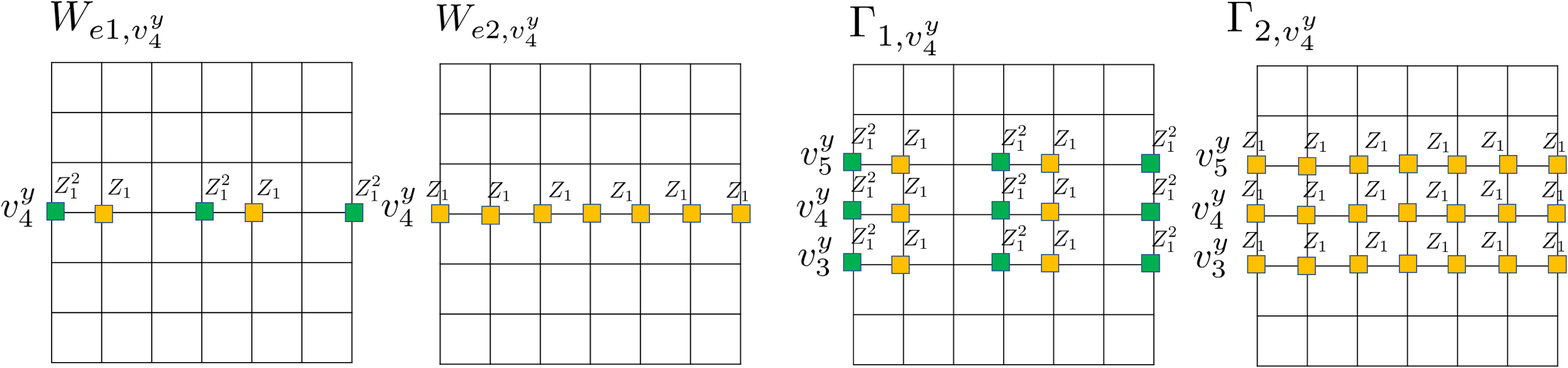}
         \caption{}\label{loops}
             \end{subfigure} \\
        \begin{subfigure}[h]{0.46\textwidth}
    \includegraphics[width=\textwidth]{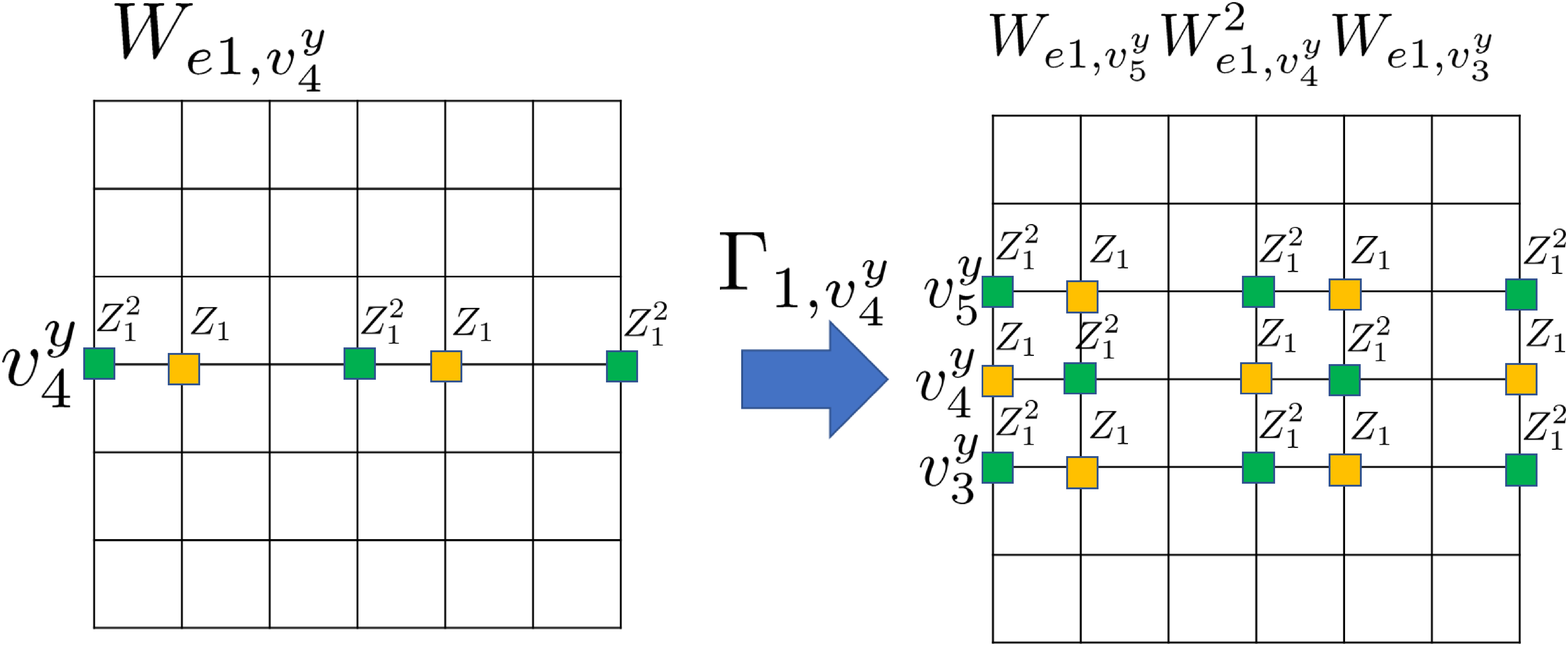}
         \caption{}\label{df1}
             \end{subfigure} 
                 \begin{subfigure}[h]{0.46\textwidth}
    \includegraphics[width=\textwidth]{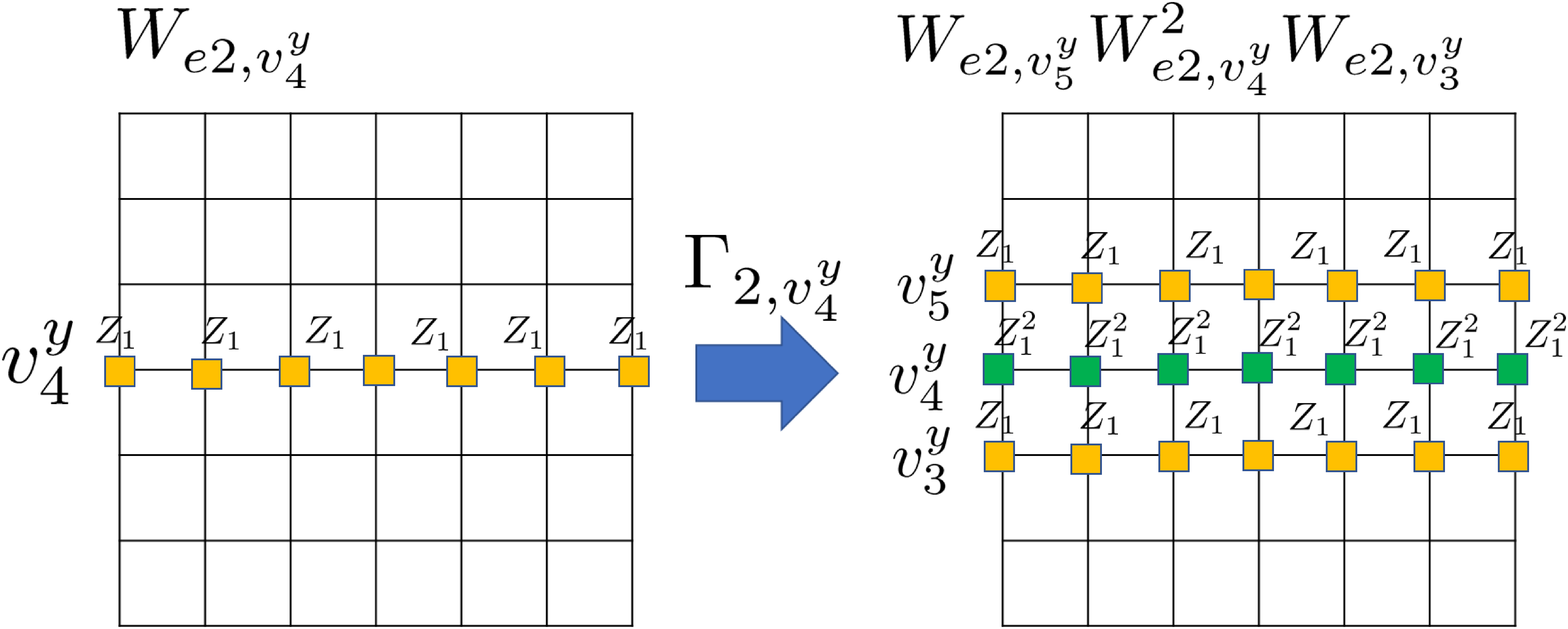}
         \caption{}\label{df2}
             \end{subfigure} 

                  \caption{ Closed loops of the electric charge in the case of $G_x\boxtimes G_y=C_6\boxtimes C_6$ and $N=3$. The periodic boundary condition is imposed so left and right edges as well as top and bottom edges are identified. 
                  (a)~(left two) Two closed loops of the electric charge in the horizontal direction
                   at $\vy{4}$, corresponding to~\eqref{lp1}. (right two) Sets of operators $P_{(v^x_i,\vy{j})}$ defined in~\eqref{haha} with which the closed loops are deformed. 
             (b)(c) Deformation of the loops in accordance with~\eqref{kk}. }
 \end{center}
 \end{figure}

Defining 
\begin{eqnarray}
 W_{e1,v^y_j}&=&\prod_{i=1}^{n_x}Z^{(\bm{\Lambda}_{1,v^y_j})_i}_{1,\vv{i}{j}}
  = Z^2_{1,\vv{1}{j}} Z_{1,\vv{2}{j}} Z^2_{1,\vv{4}{j}}Z_{1,\vv{5}{j}} ,\nonumber\\
   W_{e2,v^y_j}&=&\prod_{i=1}^{n_x}Z^{(\bm{\Lambda}_{2,v^y_j})_i}_{1,\vv{i}{j}} = 
  Z_{1,\vv{1}{j}} Z_{1,\vv{2}{j}}Z_{1,\vv{3}{j}} Z_{1,\vv{4}{j}}Z_{1,\vv{5}{j}} Z_{1,\vv{6}{j}},
  \label{lp1}
\end{eqnarray}
the closed loop of the electric charge at $v_j^y$, $W_{e,v^y_j}$ is generated by these two terms, i.e., $W_{e,v^y_j}= W_{e1,v^y_j}^{\alpha_{1,\vy{j}}}W_{e2,v^y_j}^{\alpha_{2,\vy{j}}}$. We depict these two loops~\eqref{lp1} in Fig.~\ref{loops}.
\par
Now we deform the loops in the vertical direction. Corresponding to the two vectors $\bm{\Lambda}_{1,v^y_j}$ and
 $\bm{\Lambda}_{2,v^y_j}$, we define the following two operators
 \begin{equation*}
     \Gamma_{1,\vy{j}}\vcentcolon=\prod_{i=1}^{n_x}P_{\vv{i}{j}}^{(\bm{\Lambda}_{1,v^y_j})_i},\;\; \Gamma_{2,\vy{j}}\vcentcolon=\prod_{i=1}^{n_x}P_{\vv{i}{j}}^{(\bm{\Lambda}_{2,v^y_j})_i}. 
 \end{equation*}
From~\eqref{term2}, these terms are rewritten as
\begin{eqnarray}
     \Gamma_{1,\vy{j}}&=& Z^2_{1,\vv{1}{j+1}} Z_{1,\vv{2}{j+1}} Z^2_{1,\vv{4}{j+1}}Z_{1,\vv{5}{j+1}}\times Z^2_{1,\vv{1}{j}} Z_{1,\vv{2}{j}} Z^2_{1,\vv{4}{j}}Z_{1,\vv{5}{j}}\nonumber\\
     &\times& Z^2_{1,\vv{1}{j-1}} Z_{1,\vv{2}{j-1}} Z^2_{1,\vv{4}{j-1}}Z_{1,\vv{5}{j-1}},\nonumber\\
      \Gamma_{2,\vy{j}}&=&(\prod_{i=1}^6Z_{1,\vv{i}{j-1}})\times(\prod_{i=1}^6Z_{1,\vv{i}{j}})\times (\prod_{i=1}^6Z_{1,\vv{i}{j+1}}), \label{haha}
      \end{eqnarray}
which are portrayed in Fig.~\ref{loops}. From~\eqref{lp1} and~\eqref{haha}, it follows that (see also Fig.~\ref{df1} and ~\ref{df2})
    \begin{equation}
      \Gamma_{1,\vy{j}}W_{e1,\vy{j}}=W_{e1,\vy{j+1}}W^2_{e1,\vy{j}}W_{e1,\vy{j-1}},\;\;   \Gamma_{2,\vy{j}}W_{e2,\vy{j}}=W_{e2,\vy{j+1}}W^2_{e2,\vy{j}}W_{e2,\vy{j-1}}.\label{kk}
    \end{equation}
We  need to evaluate the distinct configurations of the loops up to such deformation. 
\par
\begin{figure}
    \begin{center}
         
       \includegraphics[width=0.65\textwidth]{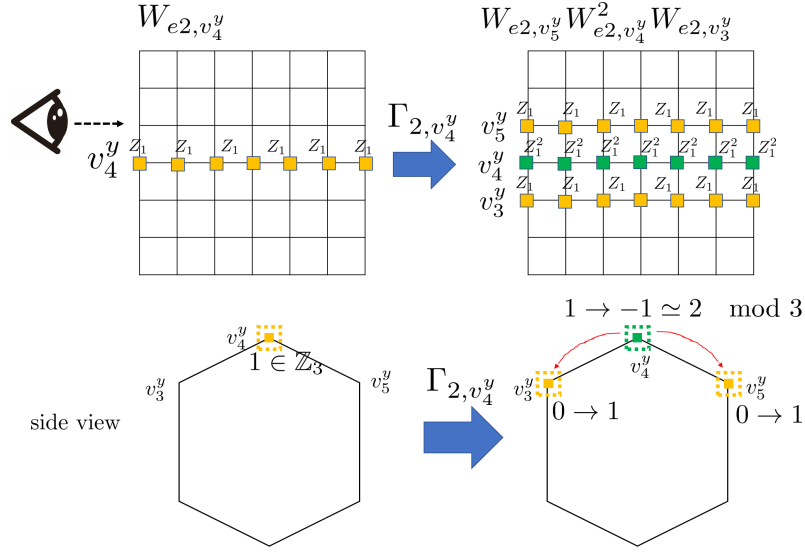}

 \end{center}
       
      \caption{Deformation of closed loops of magnetic charges in the case of $C_3\boxtimes C_3$ and $N=3$ corresponding to Fig.~\ref{df2}.
      (Top) The same figure of the deformation of the loop given in Fig.~\ref{df2}. (Bottom) The side view of Fig.~\ref{df2}, where one assigns $\mathbb{Z}_3$ number on each vertex, corresponding to the configuration of the loops. These numbers are regarded as chips located at each vertex. By applying $\Gamma_{2,\vy{4}}$, the closed loop is deformed, which corresponds to the one of the firing process where the chip at vertex $\vy{4}$ is transferred into the adjacent ones, $\vy{3}$ and $\vy{5}$ (red arrows).
 }
        \label{chip}
   \end{figure}

To this end, it is useful to draw the side view of the geometry and see how such deformation of the loops is implemented. One such example, corresponding to~\ref{df2}, is shown in Fig.~\ref{chip}. Viewing from the side, we have $G_y$, which is $C_6$ in the present case. 
At each vertex $v_j^y$, one can assign a $\mathbb{Z}_3$ number,  $\alpha_{2,\vy{j}}\in\mathbb{Z}_3$ corresponding to the closed loops of the electric charge, $W_{e2,\vy{j}}$. In Fig.~\ref{chip}, the charge $\alpha_{2,\vy{4}}=1$ is located at $v^y_4$ with charges at other vertices being absent. By applying $\Gamma_{2,\vy{4}}$, the loop is deformed, yielding the configuration on the right in Fig.~\ref{chip}: the charge located at $\vy{4}$ is decreased by two, i.e., $1\to-1\simeq 2 (\text{mod} 3)$ whereas the charge is increased by one at the adjacent vertices, $\vy{3}$ and $\vy{5}$, i.e., $0\to 1$. \par
What we have just described has an intimate relation with the \textit{chip-firing game} invented in the context of the graph theory~\cite{bjorner1991chip,biggs1999chip}. In the chip-firing game, for a given graph $G(V,E)$, 
a \textit{chip} is defined as an integer located at each vertex of the graph. Also, the process of \textit{fire} is defined as the movement of sending one chip at given vertex, say $v_0$ to each of its neighbors, which are vertices connected with $v_0$ by an edge. In the process of the fire, chip is decreased by $\text{deg}(v_0)$ at $v_0$ and at adjacent vertices the chip is increased by one. In our context, the chip introduced at each vertex corresponds to the closed loops with electric charge labeled by $\alpha_{2,\vy{j}}$, whereas the process of the fire is nothing but the deformation of the loop. 
Important distinction between the chip-firing game and our consideration is that while the chip is defined as an integer number in the chip-firing game, in our case, what corresponds to the chip is labeled by a finite group, corresponding to the charge of the fractional excitation. (In this sense, we are dealing with an anyonic analog of the chip-firing game.)\par
One of the motivations of the chip-firing game is to classify the distinct configurations of the chips up to the firing processes, and find an optimal configuration of chips. For instance, associating the chips to dollars and the vertices to money borrowers and lenders with interpreting the minus value of the chips as debt, one would be interested in finding a configuration of the chips so everyone is debt-free. (It is often referred to as the \textit{dollar game} in the context of the graph theory~\cite{biggs1999chip,BAKER2007766})
    It turns out that distinct configurations of the chips are characterized by the cokernel of the Laplacian, aka the \textit{Picard group}, $Pic(G)$~\cite{bjorner1991chip,biggs1999chip}. 
\par
To see this in a more formal fashion, we now turn to the generic cases of the 2D lattice given by~$G_x\boxtimes G_y$. As we have seen in Sec.~\ref{hori}, the closed loops of the electric charge in the horizontal direction at~$v^y_j$ are labeled by $(\alpha_{1,v^y_j},\cdots,\alpha_{m_x+1,v^y_j})\in\prod_a\mathbb
    {Z}_{\gcd(N,p_a)}\times\mathbb{Z}_N$. At $v^y_j$, the form of loops of the electric charge is given by
    \begin{equation}
       \bm{r}_{\vy{j}}=\alpha_{1,\vy{j}}\bm{\Lambda}_{1,\vy{j}}+\cdots+\alpha_{m_x,\vy{j}}\bm{\Lambda}_{m_x,\vy{
       j}}+\alpha_{m_x+1,\vy{j}}\bm{\Lambda}_{m_x+1,\vy{j}}.
    \end{equation}
    We focus on the deformation of the loop labeled by $\alpha_{a,\vy{j}}$ which we dub the loop with type $a$ $(1\leq a\leq m_x+1$).
Looking at the geometry from the side, at each vertex of graph $G_y$, $\vy{j}$, one can assign a number~$\alpha_{a,\vy{j}}$ 
associated with the configuration of the closed loops with type $a$. We define a vector $\bm{\alpha}_{a}$ as
\begin{equation}
    \bm{\alpha}_{a}=(\alpha_{a,\vy{1}},\cdots,\alpha_{a,\vy{n_y}})^T\in[\mathbb{Z}_{\gcd(N,p_a)}]^{n_y}.
\end{equation}
 (For the sake of notational simplicity, we conventionally set $p_{m_x+1}=0$ so $\bm{\alpha}_{m_x+1}\in\mathbb{Z}_N^{n_y}$.)
  Corresponding to vector $\bm{\Lambda}_{a,\vy{j}} (1\leq a\leq m_x+1)$, we define the following composite of the operators $P_{\vv{i}{j}}$
\begin{equation}
    \Gamma_{a,\vy{j}}\vcentcolon=\prod_{i=1}^{n_x}P_{\vv{i}{j}}^{(\bm{\Lambda}_{a,v^y_j})_i},\nonumber
\end{equation}
which is rewritten as
\begin{equation}
      \Gamma_{a,\vy{j}}= \biggl(\prod_{t\neq j}\biggl[\prod_{i=1}^{n_x}Z_{1,\vv{i}{l}}^{(\bm{\Lambda}_{a,v^y_j})_i}\biggr]^{l^y_{tj}}\biggr)\times \biggl[\prod_{i=1}^{n_x}Z_{1,\vv{i}{j}}^{(\bm{\Lambda}_{a,v^y_j})_i}\biggr]^{-\text{deg}^y(\vy{j})}.
\end{equation}
Using $\Gamma_{a,\vy{j}}$, we deform the loops with configuration~$\bm{\alpha}_{a}$. Suppose we deform the loop by the operator~$\Gamma_{a,\vy{1}}^{\sigma_{a,\vy{1}}}\times \cdots \Gamma_{a,\vy{n_y}}^{\sigma_{a,\vy{n_y}}}$ characterized by the vector $\bm{\sigma}_{a}=(\sigma_{a,\vy{1}},\cdots,\sigma_{a,\vy{n_y}})^T\in[\mathbb{Z}_{\gcd(N,p_a)}]^{n_y}$. Using the Laplacian of the graph $G_y$, the configuration of the deformed loop with type $a$, $\tilde{\bm{\alpha}}_a$ reads
\begin{equation}
    \tilde{\bm{\alpha}}_a=\bm{\alpha}_a-L_y\bm{\sigma}_{a}.
\end{equation}
The distinct configuration of the loop with type~$a$ up to the deformation is found to be 
\begin{equation}
   [\mathbb{Z}_{\gcd(N,p_a)}]^{n_y}/\text{im}(L_y),
\end{equation}
which is nothing but the cokernel of the Laplacian, the Picard group.

To proceed, we need to evaluate $\text{im}(L_y)$. Recalling the Laplacian is transformed into the Smith normal form
\begin{equation}
    P_yL_yQ_y=\text{diag}(u^y_1,\cdots,u^y_{n_y-1},0),
\end{equation}
we have
\begin{eqnarray}
    \text{im}(L_y)&=&L_y\mathbf{\eta},\;\forall\mathbf{\eta}\in\mathbb{Z}_{\gcd(N,p_a)}
^{n_y}\nonumber\\
&=&P_y^{-1}D_y\tilde{\mathbf{\eta}}\;(\tilde{\mathbf{\eta}}\vcentcolon=Q_y^{-1}\mathbf{\eta})\nonumber\\
&=&\text{span}(\mathbf{\pi}_1^{\prime},\mathbf{\pi}_2^{\prime},\cdots,\mathbf{\pi}_{n_y}^{\prime}).\label{qq}
\end{eqnarray}
Here, $\mathbf{\pi}_j^{\prime}$ represents the vector corresponding to the $j$-th column of $P_y^{-1}D_y$. Since $D_y$ is the diagonal with the last entry being zero, \eqref{qq} is further written as
\begin{equation}
    \text{im}(L_y)=\text{span}(u_1^y\mathbf{\pi}_1,u_2^y\mathbf{\pi}_2,\cdots,u^y_{n_y-1}\mathbf{\pi}_{n_y-1}),
    \label{decom}
\end{equation}
where $\mathbf{\pi}_j$ represents the vector which corresponds to the $j$-th column of $P_y^{-1}$. 

Now we write $\mathbf{s}_{\alpha_a}\in\mathbb{Z}_{\gcd(p_a,N)}^{n_y}/\text{im}(L_y)$ in this basis:
\begin{equation}
    \mathbf{s}_{\alpha_a}=\sum_{j=1}^{n_y}c_{a,j} \mathbf{\pi}_j\;\bigl(c_{a,j}\in \mathbb{Z}_{\gcd(p_a,N)}\bigr).
\end{equation}
From \eqref{decom}, $c_{a,j}$ is subject to (the symbol~$``\sim"$ represents identification)
\begin{equation}
    c_{a,j}\sim c_{a,j}+u^y_j\;(1\leq j\leq n_y-1)\label{con}.
\end{equation}
By definition, it also must satisfy
\begin{equation}
    c_{a,j}\sim c_{a,j}+\gcd(N,p_a)\;(1\leq j\leq n_y).\label{con2}
\end{equation}
The algebraic structure of the Picard group is determined by the number of distinct $\bm{s}$ with the two constraints~\eqref{con}\eqref{con2}.
Assuming the Smith normal form of the Laplacian $L_y$ has $m_y$ invariant factors greater than $1$, i.e., 
\begin{equation}
     D_y=\text{diag}(\underbrace{1,\cdots,1}_{n_y-1-m_y},\underbrace{q_1,\cdots,q_{m_y}}_{m_y},0)\label{snf33}
\end{equation}
then we have
\begin{equation*}
    c_{a,b^\prime}\sim c_{a,b^\prime}+1\;(1\leq b^\prime\leq n_y-1-m_y),
\end{equation*}
implying the coefficients of the first $n_y-1-m_y$ basis are trivial.

As for the coefficients $c_{a,b+n_y-1-m_y}$ $(1\leq b\leq m_y)$, they satisfy the following two conditions:
\begin{eqnarray*}
 c_{a,b+n_y-1-m_y}&\sim& c_{a,b+n_y-1-m_y}+q_b\\
  c_{a,b+n_y-1-m_y}&\sim& c_{a,b+n_y-1-m_y}+\gcd(N,p_a)\;(1\leq b\leq m_y), 
\end{eqnarray*}
from which it follows that $c_{a,b+n_y-1-m_y}\;(1\leq b\leq m_y)$ takes $\gcd(q_b,\gcd(N,p_a))=\gcd(p_a,q_b,N)$ distinct values.
Together with the fact that the last coefficient $c_{a,n_y}$ takes $\gcd(N,p_a)$ distinct values, we find that 
\begin{equation}
     \mathbf{c}_a\vcentcolon=(\underbrace{c_{a,1},\cdots,c_{a,n_y-1-m_y}}_{n_y-1-m_y},\underbrace{c_{a,n_y-m_y},\cdots,c_{a,n_y-1}}_{m_y},c_{a,n_y})^T=(\underbrace{0,\cdots,0}_{n_y-1-m_y},\underbrace{\beta_{a,1},\cdots,\beta_{a,m_y}}_{m_y},\beta_{a,m_y+1})^T \mod N\label{vecc}
\end{equation}
with $\beta_{a,b}\in \mathbb{Z}_{\text{gcd}(N,p_a,q_b)}$, $\beta_{a,m_y+1}\in\mathbb{Z}_{\text{gcd}(N,p_a)}$.
Therefore, distinct configurations of the closed loops of the charges with type $a$ are labeled by
\begin{equation}
 \mathbb{Z}_{\gcd(N,p_a,q_1)}\times \cdots\times \mathbb{Z}_{\gcd(N,p_a,q_{m_y})}\times\mathbb{Z}_{\gcd(N,p_a)}
=\prod_{b=1}^{n_y} \mathbb{Z}_{\gcd(N,p_a,q_b)}\times\mathbb{Z}_{\gcd(N,p_a)}
\end{equation}
Since
\begin{equation}
      \mathbf{s}_{\alpha_a}=\sum_{j=1}^{n_y}c_{a,j} \mathbf{\pi}_j=P_y^{-1}\mathbf{c}_a
    ,\label{hai}
\end{equation}
the explicit form of the configuration of the loops $\mathbf{s}_{\alpha_a}$ is obtained by multiplying $P_y^{-1}$ from the left in~\eqref{vecc}.
\par
Taking the deformation of the loops with all of the types into the consideration, distinct configurations of the closed loops are labeled by
\begin{equation}
 \prod_{a=1}^{m_x+1} \biggl[\prod_{b=1}^{n_y} \mathbb{Z}_{\gcd(N,p_a,q_b)}\times\mathbb{Z}_{\gcd(N,p_a)}\biggr]=
   \mathbb{Z}_N\times\prod_{a=1}^{m_x}\mathbb{Z}_{\text{gcd}(N,p_a)}\times\prod_{b=1}^{m_y}\mathbb{Z}_{\text{gcd}(N,q_b)}\times\prod_{a=1}^{m_x}\prod_{b=1}^{m_y}\mathbb{Z}_{\text{gcd}(N,p_a,q_b)}.\label{kabelss}
\end{equation}

So far, we have considered closed loops of the electric charges. Regarding the closed loops of the magnetic charges, the similar argument 
follows as the electric charges, thus they are labeled by the same quantum numbers~\eqref{kabelss}.\par
To recap the argument, we have considered distinct loops of electric and magnetic charges in our model placed on the 2D lattice $G_x\boxtimes G_y$. The GSD is obtained by counting the number of such distinct loops.
Assuming there are $m_x$ and $m_y$ invariant factors of the Laplacian $L_x$ and $L_y$ which are greater than one~[\eqref{snf22}\eqref{snf33}], indexed by $p_{a}\;(1\leq a\leq m_x)$, $q_{b}\;(1\leq b\leq m_y)$, respectively, 
we finally arrive at
\begin{equation}
  \boxed{ GSD= \bigl[N\times\prod_a\gcd(N,p_a)\times\prod_b\gcd(N,q_b)\times\prod_{a,b}\gcd(N,p_a,q_b)\bigr]^2.}\label{result}
\end{equation}
As opposed to fracton topological phases, where the GSD exhibits the sub-extensive dependence on the system size, the GSD of our model depends on $N$ and the greatest common divisor of $N$ and invariant factors of the Laplacian.
\section{Example}\label{sec5}
Our result~\eqref{result} is applicable to arbitrary connected graphs, yet it is still useful to take a closer look at the simple example of the 2D lattice, torus geometry, for an illustrative example to see how our formula works.
\subsection{Torus geometry $C_{n_x}\boxtimes C_{n_y}$}
The cycle graph $C_n$ consists of $n$ vertices placed in a cyclic order so adjacent vertices are connected by a single edge. We consider the 2D lattice constructed by the product of cycle graphs, $C_{n_x}\boxtimes C_{n_y}$.
\par
We need to find the Smith normal form of the Laplacian of the cyclic graph.
We concentrate on the transformation of $L_x$ into the Smith normal form. The Laplacian $L_x$ is described by the following $n_x\times n_x$ matrix:
\begin{equation}
    L_x=\begin{pmatrix}
2 & -1&&& -1\\
-1 & 2&-1 &&\\
&-1&2&\ddots&\\
&&\ddots&\ddots&-1\\
-1&&&-1&2
\end{pmatrix}.
\end{equation} 
Adding the first $n_x-1$ columns to the last one and doing the same procedure for the rows, 
the Laplacian 
is transformed as
\begin{equation}
  L_x  \to
\begin{pmatrix}
\tilde{L}_x&\bm{0}_{n_x-1}\\
\bm{0}_{n_x-1}^T&0\label{39}
\end{pmatrix},
\end{equation}
with
\begin{equation}
\tilde{L}_x=\begin{pmatrix}
2 & -1&&& \\
-1 & 2&-1 &&\\
&-1&2&\ddots&\\
&&\ddots&\ddots&-1\\
&&&-1&2
\end{pmatrix}_{n_x-1\times n_x-1}.
\end{equation}
Any Laplacian of the connected graph is transformed into the form~\eqref{39}, where $\tilde{L}_x$ is obtained by removing the last row and column from the Laplacian $L_x$.
We further transform $\tilde{L}_x$ as 
\begin{equation}
    \tilde{L}_x
\to
\begin{pmatrix}
1 & -2& 1&& \\
2 & -1& &&\\
&-1&2&\ddots&\\
&&\ddots&\ddots&-1\\
&&&-1&2
\end{pmatrix}\to
\begin{pmatrix}
1 & 0& 0&& \\
2 & 3&-2 &&\\
&-1&2&\ddots&\\
&&\ddots&\ddots&-1\\
&&&-1&2
\end{pmatrix}\label{23},
\end{equation}
where in the first transformation, we have exchanged the first and second rows and multiply $(-1)$ in the first rows, and in the second transformation, we have added the first column to the second one twice and subtract the first column from the third one.
By subtracting the first row fro m the second one twice, the matrix is further transformed as
\begin{equation}
   \eqref{23} \to
\begin{pmatrix}
1 & 0& 0&& \\
0 & 3&-2 &&\\
&-1&2&\ddots&\\
&&\ddots&\ddots&-1\\
&&&-1&2
\end{pmatrix}.\label{24}
\end{equation}

The last form of~\eqref{24} has a diagonal element in the $(1,1)$ entry. 
We iteratively implement the similar transformation on the sub-diagonal matrix below the $(1,1)$ entry by swapping the first and second rows of the sub-diagonal matrix followed by multiplying $(-1)$ in the first row, and adding 
the first columns and rows to or subtracting those from other 
columns and rows. Finally, one arrives at
\begin{equation}
    P_xL_xQ_x=\text{diag}(1,1,\cdots,n_x,0),\label{snf1}
\end{equation}
where matrix $P_x$ ($Q_x$) corresponds to the operations involving switching between rows (columns), negating, and adding or subtracting the rows (columns). The Smith normal form of the Laplacian $L_y$ is obtained analogously:
\begin{equation}
    P_yL_yQ_y=\text{diag}(1,1,\cdots,n_y,0)\label{snf2}
\end{equation}
\begin{figure}
    \begin{center}
       \begin{subfigure}[h]{0.46\textwidth}
  \includegraphics[width=\textwidth]{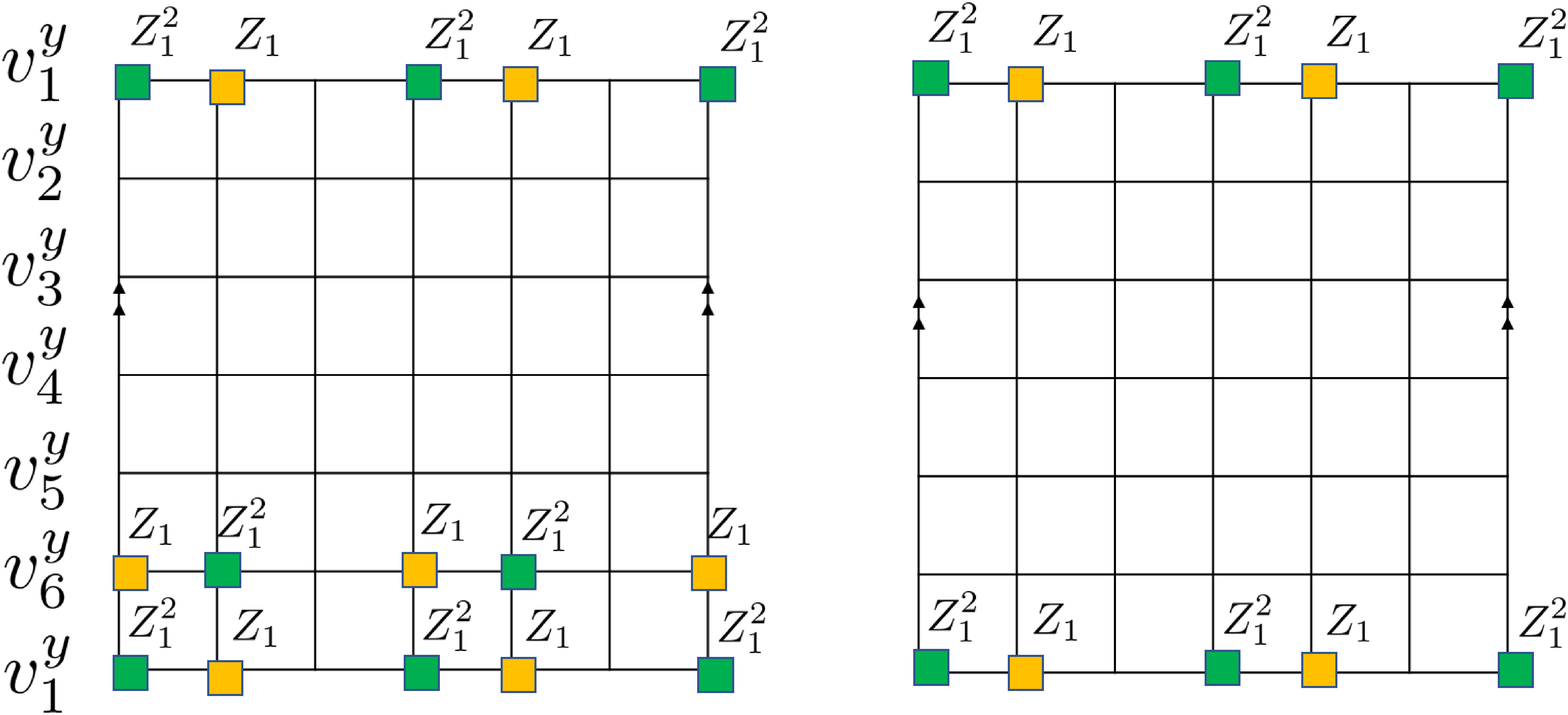}
         \caption{}\label{loop5}
             \end{subfigure} 
        \begin{subfigure}[h]{0.46\textwidth}
    \includegraphics[width=\textwidth]{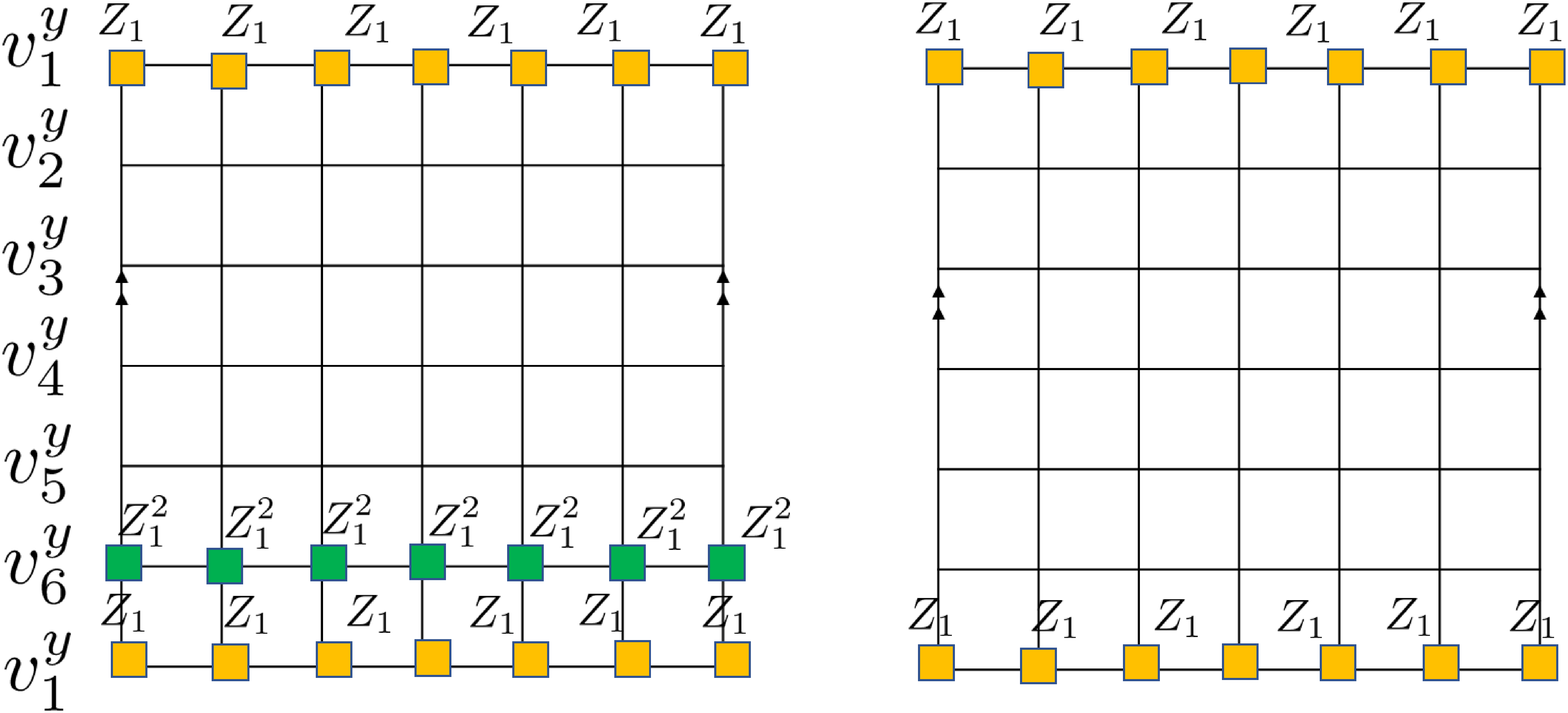}
         \caption{}\label{df5}
             \end{subfigure} \\
                 \begin{subfigure}[h]{0.76\textwidth}
    \includegraphics[width=\textwidth]{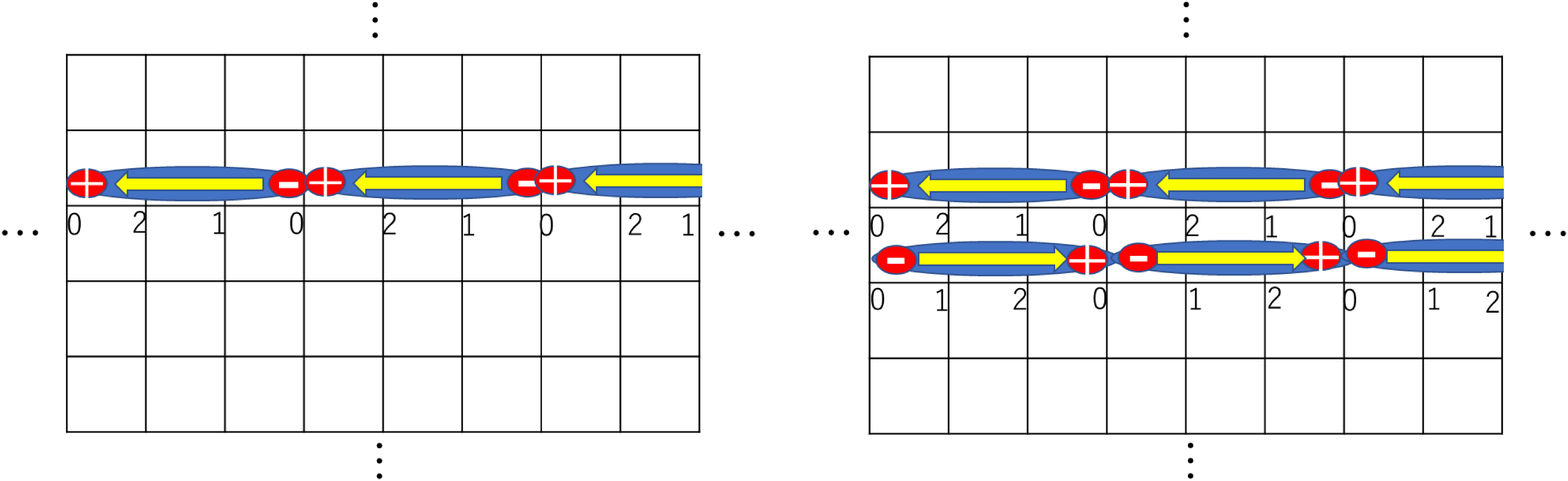}
         \caption{}\label{df55}
             \end{subfigure} 

                  \caption{ (a)[(b)] Distinct configurations of closed loops labeled by $\alpha_{1,\vy{j}}[\alpha_{2,\vy{j}}]$, corresponding to~\eqref{kini2}[\eqref{kini2dd}] in the case of $N=3$ and $n_x=n_y=6$. 
                  The periodic boundary condition is imposed in such a way that left and right edges as well as top and bottom edges are identified. 
                  (c) Left: A closed loop of dipole of the fractional charge which corresponds to~\eqref{66} in the case of $N=3$ with $n_x$ being divisible by three.
                  Regarding the pattern $``2,1,0"$, as the dipole of the fractional charges, one can make an interpretation on this loop as the arrays of such dipoles. 
                  Right: Schematic picture of the quadrupole consisting of a pair of closed loops of dipole. }
 \end{center}
 \end{figure}

\par
Based on the formula~\eqref{result}, the GSD is given by
\begin{equation}
    GSD=[N\times\gcd(N,n_x)\times\gcd(N,n_y)\times\gcd(N,n_x,n_y)]^2.
\end{equation}
By evaluating the form of the matrix $Q_x$, and referring to~\eqref{loop3} and~\eqref{con3}, the form of the closed loops of the fractional charges in the horizontal direction at $\vy{j}$ is described by
\begin{equation}
    \bm{r}_{\vy{j}}=
    N^\prime \alpha_{1,\vy{j}}\begin{pmatrix}
    n_x-1\\n_x-2\\\vdots\\1\\0
    \end{pmatrix}
    +\alpha_{2,\vy{j}}\begin{pmatrix}
    1\\1\\\vdots\\1\\1
    \end{pmatrix}\mod N, \label{kini}
\end{equation}
where $\alpha_{1,\vy{j}}\in\mathbb{Z}_{\gcd(N,n_x)}$, $\alpha_{2,\vy{j}}\in\mathbb{Z}_N$, $N^\prime=N/\gcd(N,n_x)$.
Note that $ \bm{r}_{\vy{j}}$ is an $n_x$-dimensional vector, indexed by vertices of the graph $G_x$ at $\vy{j}$.
For each loop labeled by $\alpha_{1,\vy{j}}$ and $\alpha_{2,\vy{j}}$, there are two distinct configurations up to the deformation in the $y$-direction. The distinct configurations of loops labeled by $\alpha_{1,\vy{j}}$, are described by the cokernel $\bm{s}_{\alpha_1}=[\mathbb{Z}_{\gcd(N,n_x)}]^{n_y}/im(L_y)$. By evaluating the form of $P^{-1}_y$ and referring to~\eqref{hai}, these configurations are described by
\begin{equation}
   \bm{s}_{\alpha_1}=
    \beta_{1,1}\begin{pmatrix}
    1\\
    0\\
    \vdots\\
    0\\
    -1
    \end{pmatrix}+\beta_{1,2}\begin{pmatrix}
    0\\
    0\\
    \vdots\\
    0\\
    1
    \end{pmatrix}\mod N\label{kini2},
\end{equation}
where $\beta_{1,1}=\mathbb{Z}_{\gcd(N,n_x,n_y)}$ and $\beta_{1,2}=\mathbb{Z}_{\gcd(N,n_x)}$. Note that $\bm{s}_1$ is $n_y$-dimensional vector, indexed by vertices of the graph $G_y$ and each entry corresponds to the loops going in the horizontal direction. 
We portray these two configurations in Fig.~\ref{loop5} in the case of $N=3$ and $n_x=n_y=6$.
Likewise, the distinct configurations of loops labeled by $\alpha_{2,\vy{j}}$ are given by the cokernel $\bm{s}_{\alpha_2}=[\mathbb{Z}_{N}]^{n_y}/im(L_y)$, which is found to be
\begin{equation}
   \bm{s}_{\alpha_2}=
    \beta_{2,1}\begin{pmatrix}
    1\\
    0\\
    \vdots\\
    0\\
    -1
    \end{pmatrix}+\beta_{2,2}\begin{pmatrix}
    0\\
    0\\
    \vdots\\
    0\\
    1
    \end{pmatrix}\mod N\label{kini2dd}
\end{equation}
with  $\beta_{2,1}=\mathbb{Z}_{\gcd(N,n_y)}$,  $\beta_{2,2}=\mathbb{Z}_{N}$. These configurations are depicted in Fig.~\ref{df5}.

\subsection{Physical interpretation}
In this subsection, we try to interpret the physical meaning of the configurations of the loops, especially the ones given in~\eqref{kini2}(portrayed in Fig.~\ref{loop5}), i.e., the configurations of loops labeled by $\alpha_{1,\vy{j}}$. We warn the readers that discussion presented in this subsection is schematic, yet it conveys physical intuition behind these loops.\par
 For simplicity, suppose we set $n_x$ so it is divisible by $N$, i.e., $n_x=Nd (d\in\mathbb{Z})$. Then the form of the closed loop labeled by $\alpha_{1,\vy{j}}$ which corresponds to the first term of~\eqref{kini} becomes
\begin{equation}
    (n_x-1,n_x-2,\cdots,1,0)^T=(N-1,N-2,\cdots,1,0,N-1,N-2,\cdots,1,0,\cdots,)^T\mod N, \label{pppp}
\end{equation}
where, on the right-hand side, the entry repeats the pattern $``N-1,N-2,\cdots,0"$ $d$ times. Renaming the vertex of the cyclic graph $C_{n_x}$ as $x\;(1\leq x\leq n_x)$, we define
the following vector:
\begin{equation}
    \bm{\rho}_{x}^f=(\underbrace{0,\cdots,0}_{x},1,\underbrace{0,\cdots,0}_{n_x-x-1})^T,\label{cdh}
\end{equation}
which is associated with the charge density operator of the fractional excitation, where a single fractional excitation is located at the coordinate $x$,
\eqref{pppp} is rewritten as
\begin{equation}
  \eqref{pppp}=  -\sum_{b=0}^{d-1}\biggl[\sum_{x=1}^{N}(x+b)\bm{\rho}_{x+b}^f\biggr].\label{66}
\end{equation}
This form looks familiar to us recalling the argument of the conservation of the dipole of charges in the
 higher rank Maxwell theory discussed in~\eqref{dipole}. It is tempting to regard the term inside the braket in~\eqref{66} as "the dipole of the fractional charges" as this term shows the charge monotonically decreasing as function of $x$, inducing the polarization (see also Fig.~\ref{df55}). 
Since the form of the loop~\eqref{66} repeats the pattern $``N-1,N-2,\cdots,0"$ $d$ times,
one can interpret it as the loops formed by the trajectories of the dipole of the fractional charges around in the $x$-direction, analogously to the fact that the Wilson loops are formed by the trajectory of the anyons in the topologically ordered phases.  \par
Having interpreted the form of the loop~\eqref{66} as the trajectory of the dipole of the fractional charges, now we turn to the distinct configurations of such loops up to the deformation. According to~\eqref{kini2}, any configuration of the loops is generated by two configurations. One configuration is a single loop of the dipole located at a given vertex, which corresponds to the second term of~\eqref{kini2}. Another configuration, corresponding to the first term of~\eqref{kini2}, is a pair of loops of the dipole with opposite signs located adjacent to each other in the $y$-direction, yielding a ``dipole of dipoles", which is a quadrupole of the fractional charges (Fig.~\ref{df55}). 
In summary, depending on the kernel and cokernel of the Laplacian, the phase admits closed loops of dipole or quadrupole of fractional charges, which accounts for the unusual behavior of the GSD. \par
\begin{table}[h]
  \begin{tabular}{c|c|c|c} 
  &Continuum $U(1)$ theory & Higgs phase & GSD on $G_x\boxtimes G_y$\\\hline
 conventional & Maxwell theory& $\mathbb{Z}_N$ topologically ordered phase ($\mathbb{Z}_N$ toric code) & $N^{2g_xg_y}$\\\hline
  new type & higher-rank Maxwell theory &higher-rank $\mathbb{Z}_N$ topological phase& \eqref{result}
  \end{tabular}
  \caption{Digest of this paper. We consider the topological phases obtained by gapping the higher rank Maxwell theory via Higgs mechanism on the 2D lattice $G_x\boxtimes G_y$.
  If we instead place the $\mathbb
  {Z}_N$ topologically ordered phases, obtained from the conventional Maxwell theory via Higgs mechanism, on the same lattice, the GSD is given by $N^{2g_xg_y}$, where $g_{x/y}$ represents the number of genus of the graph $G_{x/y}$, $g_{x/y}\vcentcolon=|E_{x/y}|-|V_{x/y}|+1$.}\label{tb1}
\end{table}
\section{Conclusion}\label{sec6}
Motivated by recent interest in fracton topological phases, especially in those phases on curved geometry, in this paper, we explore the geometric aspect of the unusual topological phases which admit fractional excitations with mobility constraint in a new context, graph theory. Due to the second-order derivative introduced in the higher rank Maxwell theory with which our model is defined via Higgs mechanism, the GSD of our model exhibits unusual dependence on the lattice. \par

Placing the phases on the 2D lattices beyond the regular square one, composed of two arbitrary graphs, we demonstrate that physical properties of the phases can be systematically studied by analyzing the Laplacian of the graph. We show that the fusion rules of the fractional excitations are determined by the form of the Laplacian of the graph. Furthermore, we show that the closed loops of the excitations are associated with the kernel of the Laplacian. Such loops are deformed analogously to the process of the firing in the chip-firing game, studied in the context of the graph theory. 
By making use of such analogy, we count the number of distinct configurations of the loops up to the deformation by evaluating the cokernel of the Laplacian. Based on this analysis, we derive a formula of the GSD of our phases on graphs, which depends on $N$ and invariant factors of the Laplacian. Depending on the graph, the phases admit a closed loop of dipole or quadrupole of fractional charges, which seemingly corresponds to the fact that the dipole and quadruple of charges are conserved in the higher rank Maxwell theory. Our study may contribute to understanding fracton topological phases in view of graph theory. \par
Our result is contrasted with conventional topological phases whose GSD depends on global topology of the lattice, i.e., the number of genus. For instance, if we introduce the $\mathbb
{Z}_N$ toric code, which is obtained by gapping the gauge group via Higgs mechanism in the usual Maxwell theory, and place it on the 2D lattice $G_x\boxtimes G_y$, the GSD depends on the total number of genus, thus $GSD=N^{2g_xg_y}$, where $g_{x/y}$ represents the genus of the graph $G_{x/y}$,  $g_{x/y}\vcentcolon=|E_{x/y}|-|V_{x/y}|+1$.
Such comparison is summarized in Table.~\ref{tb1}.

There are several future directions regarding the research presented in this paper. It is important to address the stability of the closed loops of fractional charges in view of quantum information as these can be utilized for logical operators. The stability can be analyzed by evaluating invariant factors of the sub-matrix of the Laplacian. It would be interesting to see whether the condition of having the stable loops is associated with other quantities of the graph such as connectivity. \par
Recently, it was proposed that the fracton topological phases can be constructed by networks of defects in topologically ordered phases~\cite{defect2018,PhysRevResearch.2.033300,PhysRevResearch.4.023258}.
It would be interesting to see how our model on graphs can be realized by the topologically ordered phases with defects. As we have seen in Sec.~\ref{2pt2}, the model with $N=2$ on the square lattice can be decomposed into copies of the toric codes. (See also~\cite{PhysRevB.97.235112,PhysRevB.100.125150}.) It would be intriguing to see whether or not our model is regarded as copies of the toric codes in the generic case of $N$ on generic lattices.
\par

In this paper, we have considered Abelian higher rank topological phases. One would naively wonder the case with non-Abelian topological phases. To study the closed loops of non-Abelian fractional charges systematically, one would consider the ``non-Abelian chip-firing game", the chip-firing game with each chip associated with non-Abelian fractional charges, which is interesting on its own right in both of graph theoretical and physical point of view. While intensive studies have been done in the case of bosonic fracton phases, much is not elucidated in the fermionic theories (and even more exotic supersymmetric theories~\cite{honda2022scalar}). Extension of our study to the fermionic cases would be another direction. 

One could
investigate other topological quantities of the model. For example, 
it would be intriguing to study entanglement
entropy of our phases on graphs and see how different it is from the case of the topologically ordered phases~\cite{ebisuBo2302ee}. 
It is well-known that in the topologically ordered phases, the total quantum dimension is related to the topological entanglement entropy~\cite{levinwen2006,preskillkitaev2006}, which is the sub-leading constant term of the entanglement entropy. 
Since the total quantum dimension crucially depends on the geometry in our model, it is worth studying to see whether such a
number enters in entanglement entropy of various geometries of subsystems.

\bibliographystyle{ieeetr}
\section*{Acknowledgement}
The author thanks Bo Han and Masazumi Honda for helpful discussion.

\bibliography{main}
\end{document}